%% file: main_MIA.tex
\documentclass[preprint,12pt,authoryear]{elsarticle}

\usepackage{multibib}
\newcites{Supp}{REFERENCES}
\usepackage{amsfonts}
\usepackage{amsmath}
\usepackage{amssymb}
\usepackage{amsthm}
\usepackage{mathtools}
\usepackage{graphicx}
\usepackage{xcolor}
\usepackage[authoryear]{natbib}
\usepackage{algorithm}
\usepackage{subcaption}
\usepackage{booktabs}
\usepackage{algorithmic}
\usepackage{caption}
\usepackage[toc,page]{appendix}
\usepackage{multirow}

\theoremstyle{plain}

\newtheorem{theorem}{Theorem}

\theoremstyle{remark}
\newtheorem{remark}{Remark}
\newtheorem{definition}{Definition}

\newcommand{\showrevisions}{0}
\newif\ifshowrevisions
\showrevisionstrue   % Set the flag to true to display revised text in blue

\newcommand{\revised}[1]{%
    \if1\showrevisions%
        \textcolor{blue}{#1}%
    \else%
        #1%
    \fi%
}
\begin{document}

\begin{frontmatter}

\title{A Deep Learning Approach to Multi-Fiber Parameter Estimation and Uncertainty Quantification in Diffusion MRI}

\author[1]{William Consagra\corref{cor1}}
\cortext[cor1]{Corresponding author: consagra@mailbox.sc.edu}
\author[2]{Lipeng Ning}
\author[2]{Yogesh Rathi}

\address[1]{Department of Statistics, University of South Carolina, Columbia, SC 29225}
\address[2]{Psychiatry Neuroimaging Laboratory, Brigham and Women’s Hospital, Harvard Medical School, Boston, MA 02215}

\begin{abstract}
Diffusion MRI (dMRI) is the primary imaging modality used to study brain microstructure \textit{in vivo}. Reliable and computationally efficient parameter inference for common dMRI biophysical models is a challenging inverse problem, due to factors such as  variable dimensionalities (reflecting the unknown number of distinct white matter fiber populations in a voxel), low signal-to-noise ratios, and non-linear forward models. These challenges have led many existing methods to use biologically implausible simplified models to stabilize estimation, for instance, assuming shared microstructure across all fiber populations within a voxel. In this work, we introduce a novel sequential method for multi-fiber parameter inference that decomposes the task into a series of manageable subproblems. These subproblems are solved using deep neural networks tailored to problem-specific structure and symmetry, and trained via simulation. The resulting inference procedure is largely amortized, enabling scalable parameter estimation and uncertainty quantification across all model parameters. Simulation studies and real imaging data analysis using the Human Connectome Project (HCP) demonstrate the advantages of our method over standard alternatives. In the case of the standard model of diffusion, our results show that under HCP-like acquisition schemes, estimates for extra-cellular parallel diffusivity are highly uncertain, while those for the intra-cellular volume fraction can be estimated with relatively high precision.
\end{abstract}


\end{frontmatter}

\input{manuscript_revision}

\section{Acknowledgements}
The authors would like to acknowledge the following grants which supported this work: R01MH119222, R01MH125860, R01MH132610, R01NS125307 and T32MH016259.

\bibliographystyle{chicago}
\bibliography{refs}

\clearpage
\pagebreak

\begin{center}
{\huge\bf SUPPLEMENTAL MATERIAL}
\end{center}

\setcounter{equation}{0}
\setcounter{figure}{0}
\setcounter{table}{0}
\setcounter{section}{0}
\setcounter{page}{1}
\makeatletter
\renewcommand{\theequation}{S\arabic{equation}}
\renewcommand{\thefigure}{S\arabic{figure}}
\renewcommand{\thetable}{S\arabic{table}}
\renewcommand{\thesection}{S\arabic{section}}
\renewcommand{\bibnumfmt}[1]{[S#1]}
\renewcommand{\citenumfont}[1]{S#1}
\renewcommand{\theequation}{S.\arabic{equation}}
\renewcommand{\thesection}{S\arabic{section}}
\renewcommand{\thesubsection}{S\arabic{section}.\arabic{subsection}}
\renewcommand{\thetable}{S\arabic{table}}
\renewcommand{\thefigure}{S\arabic{figure}}
\renewcommand{\thetheorem}{S\arabic{theorem}}
\renewcommand{\theproposition}{S\arabic{proposition}}
\renewcommand{\thelemma}{S\arabic{lemma}}
\renewcommand{\theassumption}{S\arabic{assumption}}
\renewcommand{\thealgorithm}{S\arabic{algorithm}}

\input{supplement_revision}
\bibliographystyleSupp{chicago}
\bibliographySupp{refs}

\end{document}

%% file: manuscript_revision.tex
\section{Introduction}

Diffusion magnetic resonance imaging (dMRI) is an imaging modality used extensively for studying brain structure \textit{in vivo} \citep{basser1996}. In dMRI, Pulsed Gradient Spin-Echo sequences are used to sensitize imaging signals to the local diffusion of water molecules. Due to the impact of the local neural tissue environment on this diffusion, analyzing these signals can reveal properties of the underlying brain tissue microstructure, providing a view into the cellular organization of the living human brain. These microstructural properties serve as interpretable biomarkers and are utilized in various neuroscientific tasks, such as disease diagnosis, neurosurgical planning and precision medicine \citep{walid2017,jelescu2017,novikov2019,pieri2021}. %Consequently, the development of robust statistical and computational methodologies to infer these microstructural features from the observed dMRI signals is of foundational importance.
\revised{However, the development of robust statistical and computational methodologies to infer these microstructural features from observed dMRI signals is complicated by the often ill-posed nature of the problem \citep{Kiselev2017}. Addressing these challenges requires designing methods that not only improve estimation accuracy, but also enable rigorous uncertainty quantification to assess the reliability of the estimates.} %check this for reliability definition: Essentials of Statistical Methods for Assessing Reliability and Agreement in Quantitative Imaging
\par 
Microstructural inference requires two ingredients: a model relating the tissue parameters to observed signals and an estimation procedure to recover these parameters from the observed data. In regard to the former, a standard biophysical model in dMRI is the convolution:
\begin{equation}\label{eqn:general_forward_model}
        f(\boldsymbol{p}, b) \approx \mathcal{G}[g|b,\boldsymbol{\xi}](\boldsymbol{p}) := \int_{\mathbb{S}^2}h_{\mathcal{G}}(\boldsymbol{p}^{\intercal}\boldsymbol{u}|b, \boldsymbol{\xi})g(\boldsymbol{u})d\boldsymbol{u},
\end{equation}
\citep{jian2007}, where the observed diffusion signal $f(\boldsymbol{p},b)$ is a function of two experimentally controlled acquisition parameters: the \textit{b-vector} $\boldsymbol{p}\in\mathbb{S}^2$, which is the direction of the applied magnetic field gradient, and \textit{b-value} $b\in\mathbb{R}^{+}_0$, which is a composite parameter related to the gradient strength and diffusion time. The forward map $\mathcal{G}$ is a rotationally invariant integral operator with kernel $h_{\mathcal{G}}$, parameterized by the b-value and local tissue parameters $\boldsymbol{\xi}$, and $g$ is a density function on $\mathbb{S}^2$, termed the (fiber) \textit{orientation distribution function} (ODF). The ODF describes the directional profile of coherently aligned populations of white matter fibers in a voxel, while the form of the kernel function $h_{\mathcal{G}}$ specifies the underlying biophysical model of diffusion. The approximate relation in \eqref{eqn:general_forward_model} is to highlight the various simplifying assumptions that are used in deriving the biophysical model, e.g. idealized model of the underlying tissue geometry, Gaussian phase assumption, negligible water exchange between intra and extra axonal tissue compartments,  etc.
\par 
A variety of approaches to parameter estimation under model \eqref{eqn:general_forward_model} have been proposed. When orientational inference is the primary aim, a popular class of deconvolution methods use a two-stage approach: first fixing the forward model, e.g. using a global estimate $\widehat{\boldsymbol{\xi}}$, and then solving the resulting linear inverse problem in the function space of $g$ by employing a variety of smoothness and/or sparsity promoting priors \revised{\citep{descoteaux2007,descoteaux2009, tournier2007,deslauriers2013,jeurissen2014,michailovich2011,rathi2013,sedlar2021,elaldi2021,elaldi2024,consagra2024}}. On the other end of the spectrum, a class of microstructure-focused methods target $\boldsymbol{\xi}$ while treating the orientational content as a nuisance, using the rotational invariance between $f$ and $\boldsymbol{\xi}$ implied by model \eqref{eqn:general_forward_model} to derive a non-linear regression that links the rotational invariants of the signal to the kernel parameters, with parameter estimation performed using direct optimization \citep{novikov2018} or learning based approaches \revised{\citep{reisert2017,diao2023,Jallais2024}}. Methods focusing on the joint estimation of both orientation and kernel parameters are somewhat limited. Typical approaches approximate the maximum likelihood point estimate (MLE) 
by employing standard numerical optimizers to solve the (usually non-convex) inverse problem \citep{zhang2012,panagiotaki2012,jelescu2016,harms2017}. \revised{Alternatively, general purpose MCMC-based sampling schemes for inference under a Bayesian set-up \citep{jbabdi2007,alexander2008,powell2021}, or two-stage convex reformulations via dictionary-based approaches with sparsity inducing regularization \citep{Daducci2015AMICO,YE2017288} can be applied.} 
\par 
Due to limited imaging resolution, a voxel often contains multiple distinct fiber populations, each of which can exhibit unique microstructural characteristics \citep{de2016resolving,ning2019joint}. A significant limitation of model~\eqref{eqn:general_forward_model} is that the biophysical parameters $\boldsymbol{\xi}$ are constrained to be shared between all of the fibers within a voxel. To make fiber specific inference, a natural extension of \eqref{eqn:general_forward_model} is the mixture convolution model:
\begin{equation}\label{eqn:general_mixture_forward_model}
    f(\boldsymbol{p}, b) \approx  \sum_{i=1}^n \int_{\mathbb{S}^2}w_ih_{\mathcal{G}}(\boldsymbol{p}^{\intercal}\boldsymbol{u}|b,\boldsymbol{\xi}_i)g(\boldsymbol{u}|\boldsymbol{m}_i)d\boldsymbol{u},\quad \sum_{i=1}^n w_i = 1
\end{equation}
where $\boldsymbol{\xi}_i$ are the $i$'th fibers kernel parameters, $\boldsymbol{m}_i\in\mathbb{S}^2$ is the $i$'th orientation (ODF mode), $w_i\ge 0$ is the mixture weight, and $n$ is the number of fibers passing though the imaging voxel, all of which are unknown in practice and must be estimated. Adapting the standard methods used for joint orientational and kernel parameter estimation in model~\eqref{eqn:general_forward_model} to the mixture formulation \eqref{eqn:general_mixture_forward_model} is non-trivial. The increased dimensionality of the parameter space exacerbates issues related to the non-convexity of the estimation problem, which can result in a potentially large number of local solutions and thus diminish the efficacy of standard optimization routines used for finding the MLE. \revised{Multi-fiber extensions to dictionary-based methods have been proposed \citep{Auría2015}; however, the discrete sampling of the underlying parameter space required to construct the dictionary can pose significant computational challenges, particularly when dealing with complex biophysical kernels or when incorporating desirable sparsity-inducing regularization \citep{Yap2016}. Moreover, neither standard optimization approaches nor dictionary-based methods explicitly address uncertainty quantification in their estimates.} Issues also surface in the Bayesian formulations, as the non-linear forward map can lead to a complex multimodal posterior which poses computational challenges for effective MCMC-based sampling \citep{yao2022JMLR}. Moreover, a new challenge emerges with the need to select $n$. Standard approaches to model selection, e.g. AIC/BIC \citep{tabelow2012,wong2016}, require forming the MLE for all $n$, leading to undesirable repeated model fitting. A fully Bayesian integration of the unknown $n$ necessitates an augmentation of the sampling scheme to allow for varying dimensional parameter spaces \citep{kaden20102},
%e.g. reversible-jump MCMC \citep{kaden20102}, 
which serves to further intensify the computational issues. Moreover, for $n>1$, this approach must handle the so-called ``label switching'' problem \citep{stephens2000}.
\par 
Though methods explicitly developed for full parameter inference under~\eqref{eqn:general_mixture_forward_model} are limited, we note that several related approaches based on mixture models of direct basis expansions of the diffusion signal have been developed. Specifically, \cite{tuch2002_mixturemodels} propose a mixture diffusion tensor model, with parameter estimation performed via gradient descent with multiple initialization, \cite{rathi2009} decompose the signal using a mixture Watson model and perform parameter estimation using non-linear least squares, and \cite{malcom2010} infer a mixture tensor model using an unscented Kalman filter (UKF) in the context of tractography. \cite{tabelow2012} and \cite{wong2016} propose (different) alternative reparameterizations of the mixture diffusion tensor model to improve parameter identifiability and develop specialized optimization procedures with initializations chosen via the optima of simpler approximate models. Note that all of these methods promote some form of (undesirable) repeated model fittings to avoid problems associated with non-convexity and/or for selecting $n$, save the UKF approach, which instead leverages an along-tract prior to regularize the problem. Moreover, only the UKF quantifies uncertainty in the parameter estimates, however, these are only formed during tractography and with a fixed number of fibers (and hence parameter space). 
\par 
Motivated by these shortcomings, in this work we propose a novel approach for full parameter estimation and uncertainty quantification for models of the form \eqref{eqn:general_mixture_forward_model} under standard dMRI data acquisition designs. To avoid the outlined complexities associated with simultaneously inferring all model parameters, our method adopts a sequential approach that decomposes the problem into two stages: a marginal subproblem for orientational inference and model selection, and a conditional subproblem for the remaining kernel parameters. In the first stage, a symmetry-aware deep neural network approximates the inverse mapping from signal $f$ directly to ODF $g$. This formulation effectively marginalizes out the effect of the kernel parameters, allowing simple non-parametric estimation of the number of fibers and their orientations via the maxima of $g$. In the second stage, we show that, conditional on the orientations, a model reparameterization allows the observed signal to be effectively ``demixxed'' into fiber specific features by solving a standard optimization problem. These demixxed features are then used as input to a deep conditional density estimator that infers the fiber specific kernel parameters. The deep inverse models are trained through simulations and then deployed on observed data, which amortizes many of the steps of the final inference algorithm, permitting the development of computationally scalable (re)sampling methods for uncertainty quantification for all model parameters.
\par 
The remainder of the paper is organized as follows. Section~\ref{sec:models} outlines the considered biophysical forward and measurement models, explores parameter identifiability and discusses synthetic data generation. The proposed methodology for parameter estimation and uncertainty quantification is detailed in Section~\ref{sec:method}. Section~\ref{sec:experiments} provides implementation details and experimental evaluation on both synthetic and in vivo datasets, including comparisons with relevant alternative methods. Section~\ref{sec:discussion} provides additional discussion of the results and avenues for future research. A concluding summary is provided in Section~\ref{sec:conclusion}.

\section{Models and Simulation}\label{sec:models}

\subsection{Forward Model: Mixture of Standard Models of Diffusion}\label{ssec:forward_model}
\begin{figure}
    \centering
    \includegraphics[width=\textwidth]{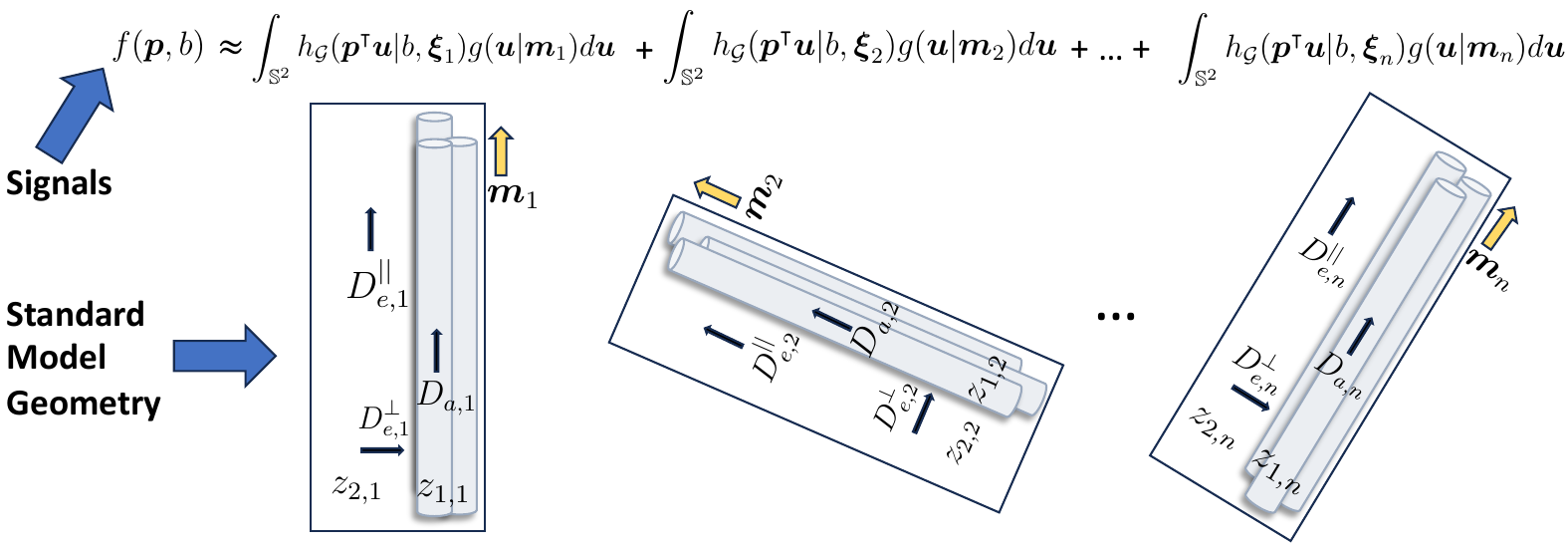}
    \caption{The standard model of diffusion \revised{in white matter} decomposes the per-voxel diffusion signal contributions into a mixture of intra and extra axonal components. Multiple distinct fiber populations can be accommodated under a mixture formulation. The observed signal can be approximated as a mixture of convolutions between the multi-compartmental kernel function and the corresponding ODF mode.}
    \label{fig:standard_model_geom}
\end{figure}
\revised{In this work, we adopt the standard model of diffusion in the white matter}, which is a unification of a variety of biophysical models that have been proposed in the literature and has become a preferred model for microstructural modeling \citep{novikov2018}. The standard model is defined by the bi-exponential kernel function 
\begin{equation}\label{eqn:standard_model_kernel}
  h_{\mathcal{G}}(\boldsymbol{p}^{\intercal}\boldsymbol{u}| b, (D_{a}, D_{e}^{||}, D_{e}^{\perp}, v)) = v\exp\left(-b D_{a}(\boldsymbol{p}^{\intercal}\boldsymbol{u})^2\right) + (1-v)\exp\left(-b D_{e}^{\perp} - b(D_{e}^{||} - D_{e}^{\perp})(\boldsymbol{p}^{\intercal}\boldsymbol{u})^2\right).
\end{equation}
The parameters $D_{a}$ and $ D_{e}^{||}$ are the intra and extra axonal diffusivity, respectively, parallel to the local white matter fiber direction, $D_{e}^{\perp}$ is the extra-axonal diffusion perpendicular to the fiber direction and $v\in [0,1]$ is the intra-axonal volume fraction. 
\par 
To model the (possibly multimodal) ODF $g$, we use a mixture of Watson distributions:
\begin{equation}\label{eqn:watson_pdf}
\begin{aligned}
    g(\boldsymbol{u}) := n^{-1}\sum_{i=1}^ng_i(\boldsymbol{u}|\boldsymbol{m}_i), \quad g_i(\boldsymbol{u}|\boldsymbol{m}_i) = C(\kappa)\text{exp}(\kappa (\boldsymbol{m}_i^{\intercal}\boldsymbol{u})^2), \quad C(\kappa) = \left[2\pi\int_0^1\exp(\kappa t^2)dt\right]^{-1},
\end{aligned}
\end{equation}
%where $I$ is the modified Bessell function of the first kind and 
where $\kappa \ge 0$ is the concentration parameter. To avoid identifiability issues with the parameters in the convolution kernel \eqref{eqn:standard_model_kernel} \revised{\citep{jelescu2016}}, we assume $\kappa$ is fixed and large. Under the large $\kappa$ assumption, we can bypass the need for direct numerical integration in the forward model, since plugging \eqref{eqn:watson_pdf} and \eqref{eqn:standard_model_kernel} into \eqref{eqn:general_mixture_forward_model} and integrating under the limit $\kappa\rightarrow\infty$, we arrive at the approximate analytic forward model 
\begin{equation}\label{eqn:forward_model}
    f(\boldsymbol{p}, b) \approx \sum_{i=1}^n \Big[z_{i,1}\exp\left(-b D_{i,a}(\boldsymbol{p}^{\intercal}\boldsymbol{m}_i)^2\right) + 
    z_{i,2}\exp\left(-b D_{i,e}^{\perp} - b(D_{i,e}^{||} - D_{i,e}^{\perp})(\boldsymbol{p}^{\intercal}\boldsymbol{m}_i)^2\right)\Big] 
\end{equation}
where $z_{i,1} = w_{i}v_i$, $z_{i,2} = w_{i}(1-v_i)$, and thus $\sum_{i=1}^nz_{i,1}+z_{i,2}=1$. With slight abuse of notation, we define the parameter $\boldsymbol{\xi}_i:=(D_{a,i}, D_{e,i}^{||}, D_{e,i}^{\perp}, z_{1,i}, z_{2,i})$.
\par 
Figure~\ref{fig:standard_model_geom} provides a depiction of the implied geometry of model~\eqref{eqn:forward_model}. The total diffusion profile is assumed to arise from a mixture of multi-compartment models, each of which models intra and extra axonal diffusion relative to the dominant orientation of the $i$th fiber, along with mixture weights $z_{1,i}, z_{2,i}$ controlling the relative contribution. The orientational contribution of the $i$th fiber population is modeled as a strongly peaked unimodal axial density functions with rotational symmetry around $\boldsymbol{m}_i$.

\subsection{Observation Model}

We assume the data is collected according to a \textit{multi-shell} sampling protocol, where the observed diffusion signal function $f_{\boldsymbol{x}}(\boldsymbol{p},b)$ at imaging voxel $\boldsymbol{x}\in\mathbb{R}^3$ is measured over an angular sampling design of $M_{l}$ gradient directions: $\boldsymbol{P}_{l,M_{l}} = (\boldsymbol{p}_{l,1}, ..., \boldsymbol{p}_{l,M_{l}})\subset \mathbb{S}^2$ on a relatively sparse set of b-values $\{b_{1},...,b_{L}\}$, referred to as shells. From here on, all modeling is done at the voxel level and so we drop the spatial index $\boldsymbol{x}$ for clarity. The protocol design can be visualized as a  collection of concentric spheres in the sampling space, where the sphere radius defines the b-shell. Figure~\ref{fig:param_sensitivity}A displays a typical sampling design employed in modern acquisition protocols, featuring $L=2$ b-shells and $M=60$ directions per shell.
\par 
In practice, the acquired data contain significant measurement error, primarily due to the thermal noise in the radio frequency coils that measure the signal. The distribution of the noise depends on several factors, including the particular acquisition and reconstruction technique, as well as any additional image processing algorithms applied \citep{fernández2016}. In this work, we consider a simple homoscedastic Gaussian measurement error model: 
\begin{equation}\label{eqn:measurement_error_model}
        s_{l,m} = f(\boldsymbol{p}_{l,m}, b_{m}) + \epsilon_{l,m} \quad m=1,...,M_{l}; \quad l=1,...,L, \quad \epsilon_{l,m}\overset{iid}{\sim}\mathcal{N}(0,\sigma_{e}^2),
\end{equation}
which is common in the literature and provides a reasonable approximation as long as the signal to noise ratio is not too low \citep{gudbjartsson1995}. Denoting $\boldsymbol{s}_{l}=(s_{l,1},...,s_{l,M})^{\intercal}$, $\boldsymbol{S} = (\boldsymbol{s}_{1}, ..., \boldsymbol{s}_{L})$, and parameters $\boldsymbol{\xi}^{(n)} = (\boldsymbol{\xi}_1, ..., \boldsymbol{\xi}_n)$, $\boldsymbol{m}^{(n)} = (\boldsymbol{m}_1, ..., \boldsymbol{m}_n)$, combining the forward model \eqref{eqn:forward_model} and measurement model \eqref{eqn:measurement_error_model} results in the observed data likelihood:
\begin{equation}\label{eqn:standard_statistical_model}
\begin{aligned}
    p(\boldsymbol{S} | \boldsymbol{\xi}^{(n)}, \boldsymbol{m}^{(n)}, n, \{(\boldsymbol{P}_{M_{l},l}, b_{l})\}_{l=1}^L)  = &\prod_{l=1}^L\prod_{m=1}^M \mathcal{N}\Big(\sum_{i=1}^n z_{i,1}\Big[\exp\left(-b_{l} D_{i,a}(\boldsymbol{p}_{m}^{\intercal}\boldsymbol{m}_i)^2\right) + \\
    &z_{i,2}\exp\left(-b_l D_{i,e}^{\perp} - b_l(D_{i,e}^{||} - D_{i,e}^{\perp})(\boldsymbol{p}_{m}^{\intercal}\boldsymbol{m}_i)^2\right)\Big], \sigma_{e}^2\Big).
\end{aligned}
\end{equation}

\subsection{Parameter Estimation and Identifiability}\label{ssec:param_identifiability}

\begin{figure}
    \centering
    \includegraphics[width=\textwidth]{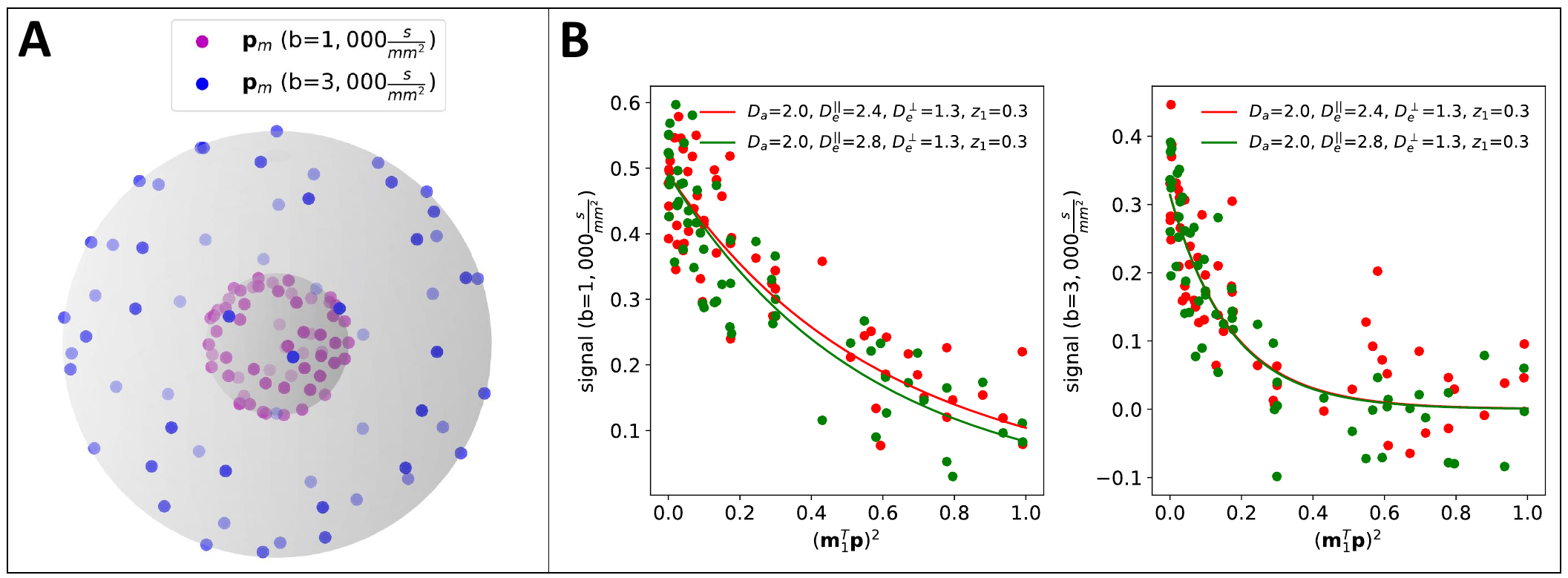}
    \caption{(A) Multi-shell sampling design. (B) Kernel decay curves for two different parameterizations, with SNR $\approx 16$. Note that, while $D_e^\parallel$ differs significantly between the two curves (red and green), the measured signals are virtually indistinguishable from each other.}
    \label{fig:param_sensitivity}
\end{figure}
The presence of the non-linear forward model in \eqref{eqn:standard_statistical_model} can render the resulting likelihood complex and multi-modal. Further complicating inference is the poor sensitivity under the multi-shell design for typical $b$-values ($\leq 4000s/\text{mm}^2$), which can result in some model parameters exhibiting poor practical statistical identifiability, i.e., very large estimation uncertainties.
%\citep{miao2011} 
Figure~\ref{fig:param_sensitivity}B provides a toy example of this issue under a simplified set-up with $n=1$ and $\boldsymbol{m}_1$ known. The red and green curves show the true signal contribution from two fiber parameterizations with different $D_{e}^{||}$. The colored points display the noisy signal observations collected under the sampling design in Figure~\ref{fig:param_sensitivity}A with a moderate SNR $\approx 16$ ($\sigma_{e}^2=0.004$). While the two parameter sets are structurally identifiable, i.e. there exists separation between the two true decay curves, the difference relative to the measurement noise renders these two situations difficult to distinguish statistically. \revised{Indeed, when $n=1$ and  $\boldsymbol{m}_1$ is known, a single b-shell design renders this a multi-exponential decay inverse problem in parameter $(\boldsymbol{m}_1^{\intercal}\boldsymbol{p})^2$—a notoriously challenging ill-posed problem with broad applications not only in medical imaging \citep{Celik2013,Martins2016,Spencer2020} but also across various fields in the physical sciences \citep{Istratov_1999}.}
\par 
To further illustrate this issue numerically, we generate 100 independent sets of signals from \eqref{eqn:standard_statistical_model} for fixed kernel parameter $\boldsymbol{\xi}_{1} = (D_{a}=1\mu \text{m}^2/\text{ms}^2, D_{e}^{||}=2\mu \text{m}^2/\text{ms}^2, D_{e}^{\perp}=1.4\mu \text{m}^2/\text{ms}^2, z_1=0.3, z_2=0.7)$ (red curves in Figure~\ref{fig:standard_model_geom}B). Fixing $n$ and $\boldsymbol{m}_1$ to their true values, we approximate the MLE of \eqref{eqn:standard_statistical_model} numerically using a standard non-linear least squares solver with multiple initializations (see description of MLE-2 in Section~\ref{ssec:competing_approaches} for details) for all 100 replicated datasets. We find the average absolute error of $D_{e}^{||}$ over the replications is $\approx 0.37$, nearly as large as the difference between the two ground truth parameter values. In more realistic settings, i.e. unknown orientations $\boldsymbol{m}_i$ and higher dimensional parameter spaces $n \ge 2$, the statistical identifiability issues using standard estimation techniques become even more pronounced (see Section~\ref{sssec:simulation_results}). 

\subsection{Simulation Model}\label{ssec:simulator}
While the complexity of function~\eqref{eqn:standard_statistical_model} may render direct likelihood-based parameter estimation difficult, combined with the availability of biologically plausible priors on the model parameters, realistic diffusion signals can be simulated cheaply. Crucially, this enables fast sampling of synthetic data from analytically intractable marginal distributions, which is then utilized to train the flexible inverse models for the sequential parameter inference procedure outlined in Section~\ref{sec:method}. %A detailed discussion of the biologically informed priors and pseudo-code for synthetic data simulation are provided in Section~\ref{ssec:sim_algos} of the supplemental materials.
\revised{A detailed discussion of the biologically informed priors for all model parameters is provided below. Pseudo-code implementing the synthetic data generation process is provided in Section~\ref{ssec:sim_algos} of the supplemental materials.}
\par\revised{The standard ranges considered biophysically plausible for the parallel diffusivities $D_{a}$ and $D_{e}^{||}$ are $[0.2, 3.0]\mu \text{m}^2/\text{ms}^2$ \citep{jelescu2016, zhang2012, panagiotaki2012}. Additional constraints on the extra-cellular perpendicular diffusivity $D_{e}^{\perp} < 0.8D_{e}^{||}$ and $D_{e}^{\perp}>0\mu \text{m}^2/\text{ms}^2$ are also imposed to avoid too much isotropy in the signals. These constraints form a polytope in the diffusivity parameter space, denoted as $\Xi := \{\boldsymbol{\xi}: \boldsymbol{A}\boldsymbol{\xi} \le \boldsymbol{d}\}$, with the form of matrix $\boldsymbol{A}$ and vector $\boldsymbol{d}$ given as}
\[
\boldsymbol{A}^\top = \begin{bmatrix}
-1 & 1 & 0 & 0 & 0 & 0 & 0 \\
0 & 0 & -1 & 1 & 0 & 0 & -1 \\
0 & 0 & 0 & 0 & -1 & 1 & 1 \\
\end{bmatrix},
\quad
\boldsymbol{d}^\top = \begin{bmatrix}
-0.2 & 3 & -0.2 & 3 & 0.0 & 2.4 & 0.0
\end{bmatrix}.
\]
\revised{We place a uniform prior over this space, independent for each $i$, which can be sampled from via rejection sampling with a box-uniform proposal. In the single fiber case, we assume a uniform prior on the $\mathbb{S}^2$ hemisphere. In the multi-fiber situation, we apply histologically motivated constraints on the crossing angle, i.e. $\text{cos}^{-1}(\boldsymbol{m}_{i}^{\intercal}\boldsymbol{m}_{j})$. Specifically, the minimum crossing angle between any two fibers must be $\ge 10$ degrees. For $n > 2$, the second smallest angle between any pair of fibers must be $\ge 30$ degrees, avoiding an undetectable ``bouquet'' type bundle configuration \citep{rheault2020common, schilling2022prevalence}.
This space can again be sampled from using rejection sampling with uniform proposals over the $n$-product space of $\mathbb{S}^2$ hemisphere. To sample the volume-fiber fractions $\{z_{i,j}\}_{i=1,j=1}^{n,2}$, we first sample from a uniform Dirichlet distribution and reject if any intra-axonal volume fraction is too small ($z_{i,1} < 0.1$). Finally, in accordance with the literature, we assume there is a max $n=3$ fibers per voxel that can be reliably detected using current acquisition parameters \citep{behrens2007probabilistic}.}

\section{Methodology}\label{sec:method}
\revised{In this section, we present our method for estimation and uncertainty quantification for all model parameters. A high-level schematic overview of the proposed methodology is presented in Figure~\ref{fig:method_overview}. Additional supporting technical details can be found in supplemental Section~\ref{sec:supp_method_details}.}
\subsection{Orientation Inversion and Model Selection}\label{ssec:orient_inference}

In this section, we outline our approach for estimating the number of fibers $n$ and their orientations $\boldsymbol{m}^{(n)}$. Due to the high concentration (large $\kappa$) assumption, these parameters can be easily calculated directly from the function $g$, i.e. by identifying the number and location of the modes. Therefore, we formulate the problem as the learning of a direct inverse mapping from the signal to $g$. A desirable consequence of this approach is that we avoid explicit modeling of $n$ and $\boldsymbol{m}^{(n)}$, and along with it the associated complexities of direct mixture model inference, e.g. variable parameter spaces, label switching. 
\par 
We first form an estimate of the diffusion signal function using basis expansion. Formally, we model 
\begin{equation}\label{eqn:spherical_basis_expansion}
    f(\boldsymbol{p},b_{l}) \approx \sum_{k=1}^K c_k(b_l)\phi_k(\boldsymbol{p}) := \boldsymbol{c}_{l}^{\intercal}\boldsymbol{\phi}(\boldsymbol{p}),
\end{equation}
where the $\phi_k:\mathbb{S}^2\mapsto\mathbb{R}$ are the real-symmetric spherical harmonic basis functions \citep{descoteaux2007}. The maximum likelihood estimates of the coefficients for each shell along with resulting signal function can be formed analytically as
\begin{equation}\label{eqn:ridge_regression}
\begin{aligned}
    &\widehat{\boldsymbol{c}}_{l} = \left[\boldsymbol{\Phi}_{M_{l}}^{\intercal}\boldsymbol{\Phi}_{M_{l}}\right]^{-1}\boldsymbol{\Phi}_{M_{l}}^{\intercal}\boldsymbol{s}_{l}, \quad  \widehat{f}_{L}(\boldsymbol{p}) := (\widehat{\boldsymbol{c}}_1^{\intercal}\boldsymbol{\phi}(\boldsymbol{p}), ...,\widehat{\boldsymbol{c}}_L^{\intercal}\boldsymbol{\phi}(\boldsymbol{p}))
\end{aligned}
\end{equation}
where $\boldsymbol{\Phi}_{M_{l}}\in\mathbb{R}^{M_{l}\times K}$ is the basis evaluation matrix with element-wise definition $\boldsymbol{\Phi}_{M_{l},m,l} = \phi_{k}(\boldsymbol{p}_{k,l})$. In low SNR or sparse angular sample (small $M_{l}$) settings, a small Laplacian-based ridge penalty can additionally be included in \eqref{eqn:ridge_regression} to reduce the variance of the estimates while still maintaining a closed form solution \citep{descoteaux2007}. This initial basis expansion is done primarily for two reasons: 1) the finite truncation level $K$ (and optional penalty term) enforce some smoothness in the signal, which is to be expected under model \eqref{eqn:standard_statistical_model} and 2) the closed form solution \eqref{eqn:ridge_regression} enables fast computation, which is crucial for the computational feasibility of the resampling based approach for uncertainty quantification outlined in Algorithm~\ref{alg:inference_algorithm}. 
\par 
Denote $\mathcal{A}$ as an inverse operator mapping the estimated signal function $\widehat{f}_L$ to $g$. We formulate an estimator of the optimal inverse operator as a minimizer of the \textit{Bayes risk} under the joint distribution $p(\widehat{f}_L,g)$ and $L^2(\mathbb{S}^2)$ loss. To avoid any theoretical difficulties with defining distributions over infinite dimensional function spaces, we assume that both $g$ and $\widehat{f}_L$ have been discretized over a dense spherical mesh, denoted here as $\boldsymbol{P}_{V}=(\boldsymbol{p}_1, ..., \boldsymbol{p}_{V})\subset\mathbb{S}^2$. Denote these discrete representations $\boldsymbol{g}\in\mathbb{R}^{V}$, $\widehat{\boldsymbol{f}}_L\in\mathbb{R}^{L\times V}$ and the discretized inverse mapping $\boldsymbol{\mathcal{A}}:\mathbb{R}^{L\times V}\mapsto\mathbb{R}^{V}$. The  Bayes risk of $\boldsymbol{\mathcal{A}}$ is given by:
\begin{equation}\label{eqn:bayes_risk}
    B(\boldsymbol{\mathcal{A}}(\cdot)) = \int \int\left\|\boldsymbol{g} - \boldsymbol{\mathcal{A}}(\widehat{\boldsymbol{f}}_L)\right\|_{2}^2p(\widehat{\boldsymbol{f}}_L|\boldsymbol{g})p(\boldsymbol{g})d\widehat{\boldsymbol{f}}_Ld\boldsymbol{g}.
\end{equation}
We define the optimal inverse mapping as the operator $\boldsymbol{\widehat{\mathcal{A}}}$ which minimizes \eqref{eqn:bayes_risk}.
\par 
Under the model proposed in Section~\ref{sec:models}, forming the Bayes risk in Equation~\ref{eqn:bayes_risk} requires calculating several complex marginalization integrals. However, the joint distribution $p(\widehat{\boldsymbol{f}}_L,\boldsymbol{g}) = p(\widehat{\boldsymbol{f}}_L|\boldsymbol{g})p(\boldsymbol{g})$ can be easily simulated from using Algorithm~\ref{alg:marginal_generative_model} in Section~\ref{ssec:sim_algos} of the supplemental materials. Therefore, we can approximate \eqref{eqn:bayes_risk} via the sampling-based counterpart: 
\begin{equation}\label{eqn:empirical_bayes_risk}
             B(\boldsymbol{\mathcal{A}}) \approx B_{I}(\boldsymbol{\mathcal{A}}):=\sum_{i=1}^I  \left\|\boldsymbol{g}_{i} - \boldsymbol{\mathcal{A}}(\widehat{\boldsymbol{f}}_{L,i})\right\|_{2}^2, \quad (\boldsymbol{g}_{i},\widehat{\boldsymbol{f}}_{L,i}) \sim p(\boldsymbol{g})p(\widehat{\boldsymbol{f}}_L|\boldsymbol{g}) := p(\widehat{\boldsymbol{f}}_L,\boldsymbol{g}).
\end{equation}
\par 
For computation, we parameterize $\boldsymbol{\mathcal{A}}$ via some model class with parameters $\psi \in \Psi$ and aim to minimize:
\begin{equation}\label{eqn:bayes_risk_minimization}
             \widehat{\psi} = \min_{\psi \in \Psi} B_{I}(\boldsymbol{\mathcal{A}}_\psi(\cdot)).
\end{equation}
It remains to be specified a reasonable model class for representing the inverse operator. Note that $\boldsymbol{\mathcal{A}}_{\psi}$ must be flexible enough to map between two very high-dimensional spaces as well as able to capture complex dependencies between the input and output. However, the structure of the model class should not be too flexible, as it can be shown that the optimal $\mathcal{A}$ obeys the following symmetry:
\begin{theorem}\label{thm:rotational_equivaraince}
   Assume the operator $\boldsymbol{\widehat{\mathcal{A}}}$ which minimizes the Bayes risk \eqref{eqn:bayes_risk} exists and is unique. Then $\boldsymbol{\widehat{\mathcal{A}}}$ is rotationally equivariant.
\end{theorem}
For additional technical details and a proof of Theorem~\ref{thm:rotational_equivaraince}, see supplemental Section~\ref{ssec:rot_equivariane_proof}.
Coupling the desire for flexibility while maintaining the symmetry property in Theorem~\ref{thm:rotational_equivaraince}, a natural parameterization of $\boldsymbol{\mathcal{A}}_{\psi}$ is via a deep neural network that is equivariant to $\mathbb{SO}(3)$. While a variety of such architectures have been proposed in the literature \citep{esteves17_learn_so_equiv_repres_with_spher_cnns,cohen2018,cobb2021efficient}, in this work, we utilize the rotationally equivariant spherical U-net architecture \citep{ronneberger2015} developed in \cite{elaldi2021}. The solution to \eqref{eqn:bayes_risk_minimization} is then approximated via stochastic gradient descent. 

\subsection{Kernel Inversion}\label{ssec:kernel_inversion}
\begin{figure}
    \centering
    \includegraphics[width=\textwidth]{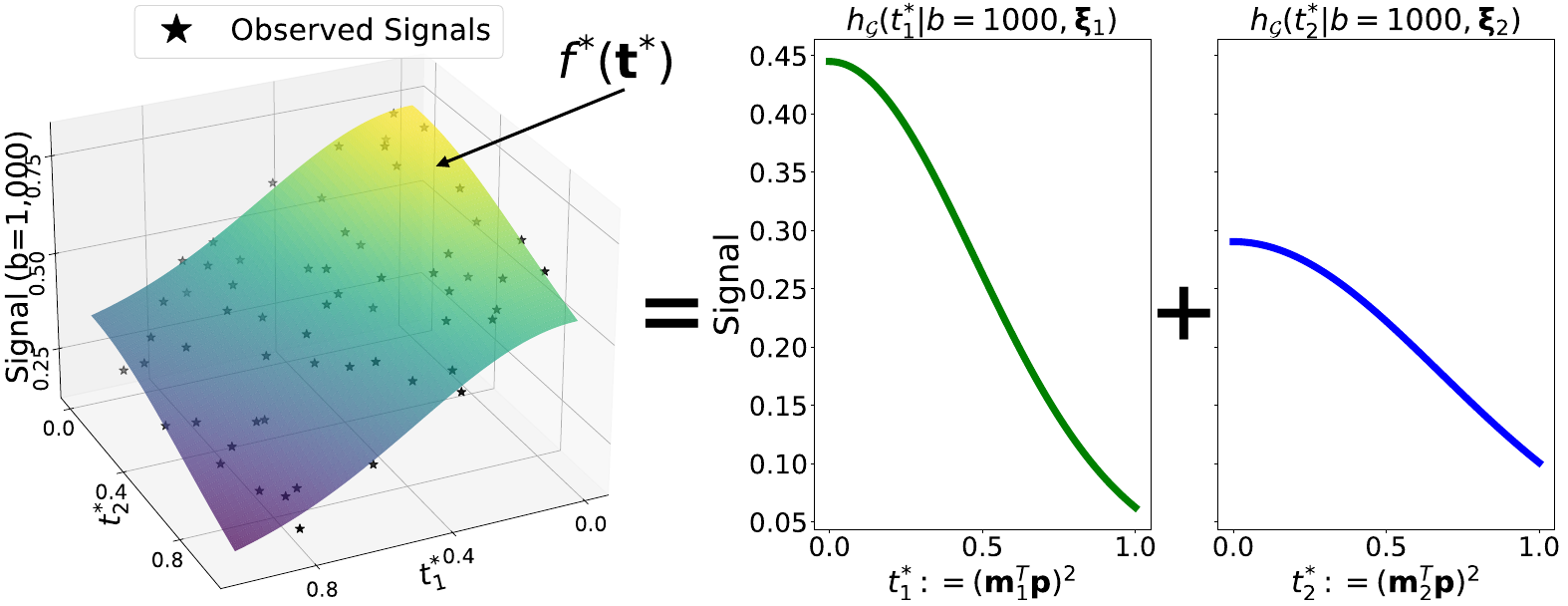}
    \caption{(Left plot) True signal function (colored surface) and noisy observations at the observed $\boldsymbol{t}_{l,m}^{*}$'s (black stars). (Right plots) True signal decomposes additively over the 1D marginal kernel decay curves.}
    \label{fig:signal_demixxing}
\end{figure}
We now outline our approach to inferring the kernel parameters $\boldsymbol{\xi}^{(n)}$, developed conditionally on the number of fibers and orientations. To motivate our approach, we first note a simple but useful re-parametrization of the signal model. Letting $t^{*}_{i} = (\boldsymbol{p}^{\intercal}\boldsymbol{m}_i)^2$, conditional on $n,\boldsymbol{m}^{(n)}$, \eqref{eqn:forward_model} can be rewritten as:
\begin{equation}
\label{eqn:general_mixture_forward_model_reparam}
    f^{*}(\boldsymbol{t}^{*},b) | n,\boldsymbol{m}^{(n)}\approx \sum_{i=1}^n h_{\mathcal{G}}(t_{i}^{*}|b,\boldsymbol{\xi}_i), \quad \boldsymbol{t}^{*} := (t_{1}^{*}, ...,t_{n}^{*}) \in [0,1]^n.
\end{equation}
For a visual representation of this model, refer to Figure~\ref{fig:signal_demixxing}. Model \eqref{eqn:general_mixture_forward_model_reparam} is recognized as an additive model in the transformed coordinate $\boldsymbol{t}^{*}$. This structure has received extensive treatment in the statistical literature \citep{buja1989, ruppert2006} and a variety of algorithms have been proposed to estimate the unknown marginal functions, i.e. the $h_{\mathcal{G}}^{(i)}:=h_{\mathcal{G}}(t_{i}^{*}|b,\boldsymbol{\xi}_i)$. As $\boldsymbol{\xi}_i$ effects the observed signals only through the $i$'th kernel decay function, it is practical to approximate our inference of the former using only our estimates of the latter, i.e., assuming \revised{approximate} conditional independence with the remaining marginal functions $j\neq i$. This approximation is beneficial computationally, as it fixes the dimension of the parameter space that needs to be modeled explicitly and helps to avoid potential issues arising from label switching. Motivated by these observations, the reminder of this section outlines our two stage estimation approach in which we first ``demixx'' the signal by estimating (a transformation of) the functions $\{h_{\mathcal{G}}^{(i)}(\cdot)\}_{i=1}^n$, and then infer $\boldsymbol{\xi}_i$ using only the corresponding function.

\subsubsection{Signal Demixxing}\label{ssec:signal_demix}
Ignoring the constraints imposed by their parametric dependence on $\boldsymbol{\xi}_i$, we first note that the functions $h_{\mathcal{G}}^{(i)}$ under model \eqref{eqn:general_mixture_forward_model_reparam} are not identifiable. This can be easily demonstrated since for any $e > 0$
$$
\begin{aligned}
    \sum_{i=1}^nh_{\mathcal{G}}(t_{i}^{*}|b,\boldsymbol{\xi}_i) &= h_{\mathcal{G}}(t_{1}^{*}|b,\boldsymbol{\xi}_1) - e + \sum_{i=2}^{n-1}h_{\mathcal{G}}(t_{i}^{*}|b,\boldsymbol{\xi}_i) + h_{\mathcal{G}}(t_{n}^{*}|b,\boldsymbol{\xi}_n) + e := \sum_{i=1}^{n}\tilde{h}_{\mathcal{G}}(t_{i}^{*}|b,\boldsymbol{\xi}_i)
\end{aligned}
$$
where 
$$
\tilde{h}_{\mathcal{G}}(t_{i}^{*}|b,\boldsymbol{\xi}_i) = \begin{cases}
   h_{\mathcal{G}}(t_{i}^{*}|b,\boldsymbol{\xi}_i) & \text{if } i = 2,...,n-1 \\
   h_{\mathcal{G}}(t_{i}^{*}|b,\boldsymbol{\xi}_i) - e & i = 1 \\
   h_{\mathcal{G}}(t_{i}^{*}|b,\boldsymbol{\xi}_i) + e & i = n. 
\end{cases}
$$
This lack of identifiability can be avoided if we consider the centered functions:
\begin{equation}\label{eqn:centered_kernel_functions}
    \bar{h}_{\mathcal{G}}^{(i,l)}(t_{i}^{*}) := h_{\mathcal{G}}(t_{i}^{*}|b_{l},\boldsymbol{\xi}_i) - \int_{0}^1h_{\mathcal{G}}(t|b_{l},\boldsymbol{\xi}_i)dt, \text{ for all }i=1,...,n,
\end{equation} 
and re-write \eqref{eqn:general_mixture_forward_model_reparam} as 
\begin{equation}\label{eqn:general_mixture_forward_model_reparam_identifiable}
    f^{*}(\boldsymbol{t}^{*},b_{l}) \approx \mu_{l} + \sum_{i=1}^n \bar{h}_{\mathcal{G}}^{(i,l)}(t_{i}^{*}), \quad \mu_{l} := \sum_{i=1}^n\int_{0}^1h_{\mathcal{G}}(t|b_{l},\boldsymbol{\xi}_i)dt.
\end{equation}
The centered functions $\bar{h}_{\mathcal{G}}^{(i,l)}$ retain the same desirable property of being independent of all $\boldsymbol{\xi}_j$ $j\neq i$. 
\par 
To obtain a finite dimensional representation for computation, we discretize the centered functions using rank $J$ cubic splines, denoted as $\boldsymbol{\gamma}(t) = (\gamma_1(t), ...,\gamma_J(t))$ and hence we model $\bar{h}_{\mathcal{G}}^{(i,l)}(t_{i}^{*}) \approx \boldsymbol{\gamma}(t_{i}^{*})^{\intercal}\boldsymbol{a}^{(i,j)}$ for unknown coefficients $\boldsymbol{a}^{(i,j)}\in\mathbb{R}^{J}$.
The reparameterization \eqref{eqn:general_mixture_forward_model_reparam_identifiable} does not change the measurement error model \eqref{eqn:measurement_error_model}, and hence we form the maximum likelihood estimates of the model parameters \eqref{eqn:general_mixture_forward_model_reparam_identifiable} via standard $l2$-minimization. Specifically, define
$\boldsymbol{t}_{l,m}^{*} := ((\boldsymbol{p}_{l,m}^\intercal\boldsymbol{m}_1)^2, ..., (\boldsymbol{p}_{l,m}^\intercal\boldsymbol{m}_n)^2)$, basis evaluation matrix $\boldsymbol{\Gamma}^{(i,l)}\in\mathbb{R}^{M\times J}$ by the element-wise definition $\boldsymbol{\Gamma}_{mj}^{(i,l)} = \gamma_j(t^{*}_{l,m,i})$, and let $\boldsymbol{1}_{M}\in\mathbb{R}^{M}$ be the column vector of ones.
The coefficients $\boldsymbol{a}^{(i,l)}$ are approximated by solving the set of optimization problems for $l=1,...,L$:
\begin{equation}\label{eqn:SCAM}
\begin{aligned}
   \widehat{\mu}_{l},\widehat{\boldsymbol{a}}^{(1,l)},...,\widehat{\boldsymbol{a}}^{(n,l)} = &\min_{\mu_{l},\boldsymbol{a}^{(1,l)},...,\boldsymbol{a}^{(n,l)}}\left\|\boldsymbol{s}_{l} - \left[\boldsymbol{1}_{M}\mu_{l} + \sum_{i=1}^n\boldsymbol{\Gamma}^{(i,l)}\boldsymbol{a}^{(i,l)}\right]\right\|_2^2 \\
   &\text{s.t.}  \quad  \boldsymbol{1}_{M}^{\intercal}\boldsymbol{\Gamma}^{(i,l)}\boldsymbol{a}^{(i,l)} = 0, \text{ for }i = 1, ..., n, \\
\end{aligned}
\end{equation}
where the constraint (approximately) enforces \revised{the integration-to-zero condition in Equation }\eqref{eqn:centered_kernel_functions}.  The function estimates are then formed as $\widehat{\bar{h}}_{\mathcal{G}}^{(i,l)}(t_{i}^{*}) = \boldsymbol{\gamma}(t_{i}^{*})^{\intercal}\widehat{\boldsymbol{a}}^{(i,l)}$, and the final estimated vector-valued function is denoted as 
$\widehat{\boldsymbol{\bar{h}}}_{\mathcal{G}}^{(i)}(t_i^{*}) := (\widehat{\bar{h}}_{\mathcal{G}}^{(i,1)}(t_i^{*}), ..., \widehat{\bar{h}}_{\mathcal{G}}^{(i,L)}(t_i^{*}))$.
\par 
There exist a variety of algorithms and parameterizations for solving the optimization problem \eqref{eqn:SCAM}. Noting that \eqref{eqn:centered_kernel_functions} are monotonic decreasing in $t_{i}^{*}$, to reduce bias we integrate this feature into the estimation by enforcing the cubic b-spline to also obey this shape constraint. This can be accomplished by reparameterizing the coefficients $\boldsymbol{a}^{(l,i)}$ using the formulation outlined in \cite{pya2015} (see supplemental Section~\ref{ssec:demix_algo} for more details). For estimation, we utilize the Newton's method optimizer to approximate the solution to \eqref{eqn:SCAM}, proposed in the same reference. \revised{For supporting details on additive model based demixxing, see supplemental Section~\ref{ssec:GAM_signal_demix}}.

\subsubsection{Posterior Approximation using Deep Neural Network}
After demixxing via \eqref{eqn:SCAM}, the problem is reduced to inferring $\boldsymbol{\xi}_i$ from the functions $\widehat{\boldsymbol{\bar{h}}}_{\mathcal{G}}^{(i)}$, for each $i=1,...,n$. This problem is ill-posed in the sense that small discrepancies between the true and estimated curves, e.g., resulting from measurements noise, finite angular samples $M$, etc., can cause different regions of the parameters space $\boldsymbol{\xi}$ to be poorly identifiable (see Figure~\ref{fig:standard_model_geom}B). This situation renders a single point estimate of $\boldsymbol{\xi}_i$ insufficient. Instead, we'd like to account for this uncertainty and characterize a set of plausible solutions.
\par 
Given these considerations, a reasonable object to base our inferences off of is the conditional (posterior) distribution. Recalling the conditional independence assumption, i.e. justified by the observation that the true centered functions $\{\bar{h}_{\mathcal{G}}^{(j,l)}\}_{j\neq i}^n$, for $l=1,...,L$, do not depend on $\boldsymbol{\xi}_i$, the object of interest is given by
\begin{equation}\label{eqn:kernel_posterior}
   p(\boldsymbol{\xi}^{(n)}|\widehat{\boldsymbol{\bar{h}}}_{\mathcal{G}}^{(1)},...,\widehat{\boldsymbol{\bar{h}}}_{\mathcal{G}}^{(n)}) \approx \prod_{i=1}^np(\boldsymbol{\xi}_i|\widehat{\boldsymbol{\bar{h}}}_{\mathcal{G}}^{(i)},n).
\end{equation}
By \eqref{eqn:kernel_posterior}, we need only consider the marginal posteriors, which by Bayes theorem may be written as
\begin{equation}\label{eqn:marginal_posterior}
p(\boldsymbol{\xi}_i|\widehat{\boldsymbol{\bar{h}}}_{\mathcal{G}}^{(i)},n) \propto p(\widehat{\boldsymbol{\bar{h}}}_{\mathcal{G}}^{(i)}|\boldsymbol{\xi}_i,n)p(\boldsymbol{\xi}_i|n).
\end{equation}
However, due to the $\widehat{\boldsymbol{\bar{h}}}_{\mathcal{G}}^{(i)}$ being defined as solutions to the optimization problem \eqref{eqn:SCAM}, we again run into the issue of an intractable likelihood function, prohibiting many common approaches to inference which require its evaluation. To avoid this issue, we propose a variational approach, where we approximate the posterior by training a flexible conditional density estimator using samples from  
$p(\boldsymbol{\xi}_i,\widehat{\boldsymbol{\bar{h}}}_{\mathcal{G}}^{(i)}|n)$. Specifically, letting $p_{\eta}(\cdot|\cdot)$ denote some flexible class of conditional density estimators, we propose the following optimization problem:
\begin{equation}\label{eqn:amortized_posterior_optimization}
    \widehat{\eta} = \max_{\eta}\mathbb{E}_{p(\boldsymbol{\xi}_i,\widehat{\boldsymbol{\bar{h}}}_{\mathcal{G}}^{(i)}|n)}\left[ \log p_{\eta}(\boldsymbol{\xi}_i|\widehat{\boldsymbol{\bar{h}}}_{\mathcal{G}}^{(i)},n)\right],
\end{equation}
where the expectation is approximated using samples via Algorithm~\ref{alg:signal_demixx_marginal_model} in supplemental Section~\ref{ssec:sim_algos} and $ \widehat{\eta}$ is formed via stochastic gradient ascent on the resulting empirical objective. 
\par It can be shown that the true posterior is a solution to the optimization problem in \eqref{eqn:amortized_posterior_optimization} \citep{papamakarios2016}. Hence, given a flexible enough model class, we can expect $p_{\widehat{\eta}}$ to be a good approximate model to base our inferences off. We parameterize $p_{\eta}$ using a mixture density network (MDN) \citep{bishop1994}, that is, a Gaussian mixture model with weights, means and covariances parameterized by a deep neural network. MDN's inherit the universal (density) approximation property from Gaussian mixture models 
\citep{GoodBengCour16}, making them an attractive choice for flexible conditional density modeling. The joint posterior is then approximated as the product of the approximate marginals:
\begin{equation}\label{eqn:appromate_posterior}
  \widehat{p}(\boldsymbol{\xi}^{(n)}|\widehat{\boldsymbol{\bar{h}}}_{\mathcal{G}}^{(1)},...,\widehat{\boldsymbol{\bar{h}}}_{\mathcal{G}}^{(n)}) \approx \prod_{i=1}^np_{\widehat{\eta}}(\boldsymbol{\xi}_i|\widehat{\boldsymbol{\bar{h}}}_{\mathcal{G}}^{(i)},n).
\end{equation}
Note that the MDN used to parameterize $p_{\eta}$ does not necessarily respect the constraints of $\boldsymbol{\xi}_i$. This can be handled practically by including a final rejection sampling step to impose the constraints on the diffusivities, followed by a projection onto the constraint set: $\sum_{i=1}^n\sum_{j=1}^2z_{i,j}=1$.

\subsection{Full Parameter  Inference}\label{ssec:estimation_and_uq}
Given the models $\boldsymbol{\mathcal{A}}_{\widehat{\psi}}$, $p_{\widehat{\eta}}$ trained using the approaches outlined in Section~\ref{ssec:orient_inference} and \ref{ssec:kernel_inversion}, 
Algorithm~\ref{alg:inference_algorithm} outlines our procedure for parameter estimation and uncertainty quantification from the observed diffusion data $\{(\boldsymbol{s}_{l}, \boldsymbol{P}_{M_{l},l}, b_{l})\}_{l=1}^L$. The algorithm takes as additional input sampling sizes $Q$ and $B$ and an estimate of measurement error variance $\widehat{\sigma}_{e}^2$. This latter quantity can be calculated in several ways. In this work, we adopt the estimator from \cite{consagra2022}, which is formed using the $b=0$ images as follows: the per-voxel mean $b=0$ signal is calculated and used to normalize each of the $b=0$ images. The per-voxel empirical variance is then computed using these normalized images and then averaged over the voxels. Under the measurement model \eqref{eqn:standard_statistical_model}, it can be shown this is an asymptotically unbiased estimator of $\sigma_{e}^2$ \citep{consagra2022}.
\begin{algorithm}[ht!]
\caption{Estimation and Uncertainty Quantification}
\label{alg:inference_algorithm}
\begin{algorithmic}[1]
\STATE \textbf{Input}: Observed data $\{(\boldsymbol{s}_{l}, \boldsymbol{P}_{M_{l},l}, b_{l})\}_{l=1}^L$, trained model $\boldsymbol{\mathcal{A}}_{\widehat{\psi}}$, $p_{\widehat{\eta}}$ estimated measurement error variance $\widehat{\sigma}_{e}^2$, number of samples $Q$, number of bootstrap replicates $B$
\STATE Form pilot estimates $\widehat{\boldsymbol{c}}_{1}, ..., \widehat{\boldsymbol{c}}_L$, $\widehat{f}_L$ via \eqref{eqn:ridge_regression} using $\boldsymbol{s}_{1}, ..., \boldsymbol{s}_{L}$
\STATE Compute and store $\widehat{\boldsymbol{f}}_l = \boldsymbol{\Phi}_{M_{l}}\widehat{\boldsymbol{c}}_{l}$, $l=1,...,L$
\STATE Estimate $\widehat{\boldsymbol{g}} = \boldsymbol{\mathcal{A}}_{\widehat{\psi}}(\widehat{f}_{L})$
and set $\widehat{\boldsymbol{m}}_1, ..., \widehat{\boldsymbol{m}}_{\widehat{n}} 
\gets \text{modes}(\widehat{\boldsymbol{g}})$
\FOR{b = 1,...,B}
\STATE Sample $\boldsymbol{\epsilon}^{(b)}_1, ..., \boldsymbol{\epsilon}^{(b)}_L \overset{iid}{\sim} \mathcal{N}(0,\widehat{\sigma}_{e}^2\boldsymbol{I})$
\STATE Sample bootstrapped signals $\boldsymbol{s}_{l}^{(b)} = \widehat{\boldsymbol{f}}_l + \boldsymbol{\epsilon}^{(b)}_l$ for $l=1,...,L$
\STATE Form bootstrapped estimate $\widehat{f}_{L}^{(b)}$ via \eqref{eqn:ridge_regression} using $\boldsymbol{s}_{1}^{(b)}, ..., \boldsymbol{s}_{L}^{(b)}$
\STATE Estimate $\widehat{\boldsymbol{g}}^{(b)} = \boldsymbol{\mathcal{A}}_{\widehat{\psi}}(\widehat{f}_{L}^{(b)})$
\STATE $\widehat{\boldsymbol{m}}_1^{(b)}, ..., \widehat{\boldsymbol{m}}^{(b)}_{\widehat{n}_{b}} 
\gets \text{modes}(\widehat{\boldsymbol{g}}^{(b)})$
\ENDFOR
\STATE Compute $\widehat{\boldsymbol{t}}_{l,m}^{*} = ((\boldsymbol{p}_{l,m}^\intercal\widehat{\boldsymbol{m}}_1)^2, ..., (\boldsymbol{p}_{l,m}^\intercal\widehat{\boldsymbol{m}}_{\widehat{n}})^2)$ and $\widehat{\boldsymbol{\Gamma}}_{mj}^{(i,l)} = \gamma_j(\hat{t}^{*}_{l,m,i})$, for $i=1,...,n$, $l=1,...,L$, $m=1,...,M$
\STATE Estimate $\widehat{\boldsymbol{a}}^{(1,l)},...,\widehat{\boldsymbol{a}}^{(n,l)}$ via \eqref{eqn:SCAM} with $\widehat{\boldsymbol{\Gamma}}_{mj}^{(i,l)}$ and $\boldsymbol{s}_{l}$ for $l=1,...,L$, and set \\
$\widehat{\boldsymbol{\bar{h}}}_{\mathcal{G}}^{(i)}=(\boldsymbol{\gamma}(t_{i}^{*})^{\intercal}\widehat{\boldsymbol{a}}^{(i,1)}, ..., \boldsymbol{\gamma}(t_{i}^{*})^{\intercal}\widehat{\boldsymbol{a}}^{(i,L)})$
\FOR{i = 1, ...,$\widehat{n}$}
\STATE Sample $\boldsymbol{\xi}_{i,q} \sim p_{\widehat{\eta}}(\boldsymbol{\xi}_i|\widehat{\boldsymbol{\bar{h}}}_{\mathcal{G}}^{(i)},\widehat{n})$ for $q=1,...,Q$
\ENDFOR
\STATE \textbf{Return} $\{\widehat{\boldsymbol{m}}_1, ..., \widehat{\boldsymbol{m}}_{\widehat{n}}\}$;  $\mathcal{M}_{B} := \{ \widehat{\boldsymbol{m}}_1^{(b)}, ..., \widehat{\boldsymbol{m}}^{(b)}_{\widehat{n}_{b}}\}_{b=1}^B$; $\{(\boldsymbol{\xi}_{1,q}, ..., \boldsymbol{\xi}_{\widehat{n},q})\}_{q=1}^Q$
\end{algorithmic}
\end{algorithm}
\subsubsection{Estimation}
The estimation of the number of fibers $n$ and their corresponding orientations $\boldsymbol{m}^{(n)}$ is accomplished by first applying the learned inverse operator to estimate $\widehat{\boldsymbol{g}} = \boldsymbol{\mathcal{A}}_{\widehat{\psi}}(\widehat{f}_{L})$, and then identifying the number and location of all local maxima of $\widehat{\boldsymbol{g}}$. These estimates are substituted for the true unknown conditioning parameters in the optimization~\eqref{eqn:SCAM}, the solutions to which are subsequently plugged into the conditional posterior \eqref{eqn:appromate_posterior} for sampling. We consider two point estimators for the kernel parameters based on these samples. The first is the posterior mean (PM), approximated using the sample mean, which is a standard estimator that is easy to compute and interpret; however, it can be problematic in the case of multi-modal posteriors. Therefore, we also consider an approximate maximum a-posteriori (MAP) estimator, which is computed as
\begin{equation}\label{eqn:approximate_MAP}
\begin{aligned}
        &\widehat{\boldsymbol{\xi}}_1^{(MAP)},...,\widehat{\boldsymbol{\xi}}_n^{(MAP)} = \underset{\boldsymbol{\xi}^{(n)}\in \{\boldsymbol{\xi}^{(n)}_{q}\}_{q=1}^Q}{\text{argmax}} \underbrace{\prod_{l=1}^L\prod_{m=1}^{M_{l}}\mathcal{N}(\sum_{i=1}^{\widehat{n}}h_{\mathcal{G}}(\boldsymbol{p}_{m,l}^{\intercal}\widehat{\boldsymbol{m}}_i|b,\boldsymbol{\xi}_i), \widehat{\sigma}_{e}^2))}_{\propto p(\boldsymbol{\xi}^{(n)} | \widehat{\boldsymbol{m}}^{(n)}, \{(\boldsymbol{s}_{l}, \boldsymbol{P}_{M,l}, b_{l})\}_{l=1}^L,\widehat{\sigma}_{e}^2)}. \\
\end{aligned}
\end{equation}
The estimator \eqref{eqn:approximate_MAP} evaluates the conditional posterior distribution over a grid of test points formed via samples from the approximate posterior \eqref{eqn:appromate_posterior}. If the latter is a good approximation of the former, this set should reside in a high-density space, allowing a relatively small number of sample points to be used for computation, thereby avoiding the curse of dimensionality that would prohibit a naive grid search over the full multidimensional space. The properties of similar multidimensional mode estimators have been studied in the literature  \citep{abraham2003}. Note that \eqref{eqn:approximate_MAP} uses the true conditional posterior, rather than the approximated one. Though somewhat heuristic, this was done in order to help remedy any error that accumulations over the sequence of approximations.

\subsubsection{Uncertainty Quantification}
To quantify uncertainty in the orientation estimates, we use a parametric bootstrap-based procedure, outlined in steps 5-10 of  Algorithm~\ref{alg:inference_algorithm}. Using the resulting set $\mathcal{M}_{B}$, we classify the orientations in each bootstrapped sample $ \{\widehat{\boldsymbol{m}}_1^{(b)}, ..., \widehat{\boldsymbol{m}}^{(b)}_{\widehat{n}_{b}}\}$ to the closest (in terms of minimum spherical distance) estimate $\widehat{\boldsymbol{m}}_{i}$, discarding the remaining sampled orientations if $\widehat{n}_{b} > \widehat{n}$. Denote the full set of bootstrapped orientations classified to $\widehat{\boldsymbol{m}}_{i}$ as $\mathcal{M}_{i}^{B}$. We compute two per-fiber scalar uncertainty measures, the detection rate (DR) and the angular dispersion (AD): 
\begin{equation}\label{eqn:orient_uq_measures}
\begin{aligned}
    \text{DR}(\widehat{\boldsymbol{m}}_i) = \frac{|\mathcal{M}_{i}^{B}|}{B}; \qquad 
    \text{AD}(\widehat{\boldsymbol{m}}_{i}) = \text{sin}^{-1}\left(\sqrt{1 - \text{EigMax}\left(\frac{1}{|\mathcal{M}_{i}^{B}|}\sum_{\boldsymbol{m}\in \mathcal{M}_{i}^{B}}\boldsymbol{m}\boldsymbol{m}^{\intercal}\right)}\right).
\end{aligned}
\end{equation}
DR$(\widehat{\boldsymbol{m}}_{i})\in [0,1]$ quantifies how often the $i$th peak was detected. $\text{AD}(\widehat{\boldsymbol{m}}_{i})$ is bounded between $0$ and $0.94$ radians, with the lower bound occurring when all bootstrapped orientations point along the same direction (low uncertainty) and the upper bound when the directions are uniformly distributed (high uncertainty).  Crucially, all of the steps in the bootstrap iteration are very fast to compute, allowing such a resampling approach to be computationally feasible.
\par 
To quantify the (conditional) uncertainty in the kernel parameters, we use the highest density regions (HDR) of the approximated marginal posteriors, whose definition is provided as follows:
\begin{definition}
The $100(1-\alpha)\%$ Highest Density Regions (HDR) is defined as the set 
$$
R(p_{\alpha}) = \{\boldsymbol{\xi}_{i}: p_{\widehat{\eta}}(\boldsymbol{\xi}_{i}|\widehat{\boldsymbol{\bar{h}}}_{\mathcal{G}}^{(i)},n) \ge p_{\alpha} \}
$$
where $p_{\alpha}$ is the largest value such that 
$\int_{R(p_{\alpha})}  p_{\widehat{\eta}}(\boldsymbol{\xi}_{i}|\widehat{\boldsymbol{\bar{h}}}_{\mathcal{G}}^{(i)},n) \ge 1-\alpha$.
\end{definition}
\noindent{The HDR is a useful object in the present context due to its ability to accommodate multi-modal posteriors, the presence of which in our case prohibits the use of simple posterior variance estimates for summarizing uncertainty. For computational ease, we construct the HDRs marginally for each parameter of interest, i.e. so the HDRs are unions of (possibly disjoint) intervals. For details on how these were computed, see supplemental Section~\ref{ssec:HDR_computation}. 
%\citep{hyndman1996}

\begin{remark}
    A potential pitfall of Algorithm~\ref{alg:inference_algorithm} is that there may exist distinct fiber orientations that shows up in the bootstrapped samples but not in the original estimates, and hence would not be detected in our procedure. This issue could be mitigated by using the full set of bootstrapped orientations with a non-parametric clustering algorithm that also selects the number of clusters. This was not pursued here to maintain computational speed.
\end{remark}
\begin{figure}[!ht]
    \centering
    \includegraphics[width=\textwidth]{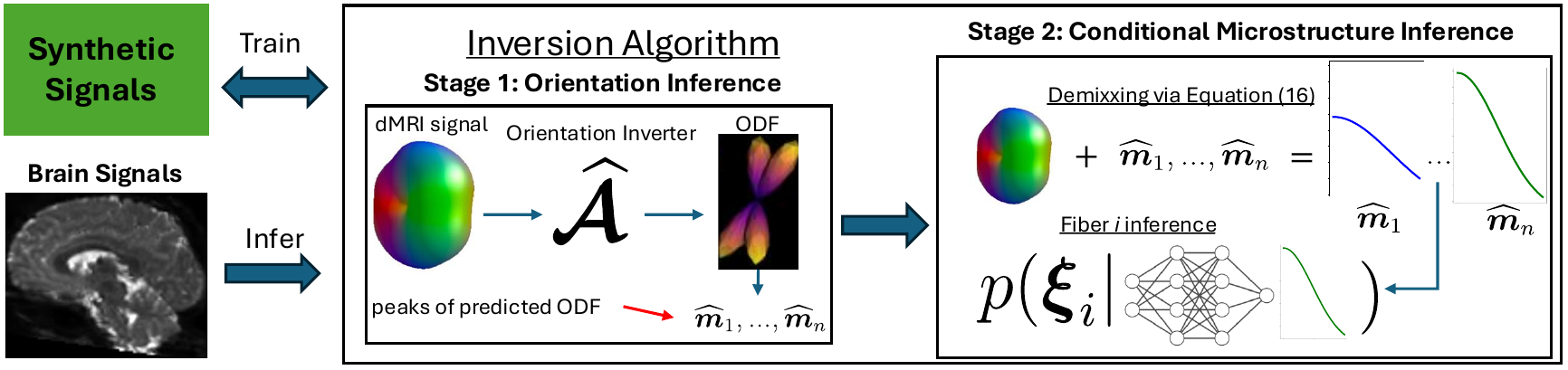}
    \caption{\revised{Outline of the methodology. The inversion algorithm is trained in two stages. In the first stage, the orientation inverter is trained to map signals, with varying kernels and measurement errors, to the underlying ODF using synthetic data generated via Algorithm~\ref{alg:marginal_generative_model} (supplemental text).
    The number and peaks of the ODF estimate the fiber orientations, which are then input, along with the signal, into the demixxing procedure by solving the optimization problem in Equation~\eqref{eqn:SCAM}. The signal curves are then used as fiber-specific summary statistics to train a deep mixture density estimator, which approximates the conditional posterior of the $i$'th fiber's microstructure. This training utilizes synthetic data generated with Algorithm~\ref{alg:signal_demixx_marginal_model} (supplemental text). After training, the inversion algorithm can be applied to real brain signals for fast parameter inference and uncertainty quantification via Algorithm~\ref{alg:inference_algorithm}.}}
    \label{fig:method_overview}
\end{figure}
\section{Experiments}\label{sec:experiments}
\subsection{Datasets}

\subsubsection{In vivo Data}
We evaluate our method using the Human Connectome Project Young Adult (HCP-YA) dataset \citep{vanessen2013}. The imaging protocol and basic preprocessing are reported in \citep{glasser_2013}. We consider a randomly selected participant (subject ID 715041). The diffusion images for this subject were obtained using a 2D spin-echo EPI sequence with a multi-shell sampling scheme, measuring signals at $M=60$ directions for b-values of 1,000, 2,000, and 3,000 $s/\text{mm}^2$, along with 12 b=0 images. The spatial  resolution was 1.25 $\text{mm}^3$. In this work, we consider the sub-design corresponding to the $b_{1}=1,000s/\text{mm}^{2}$ and $b_{2}=3,000s/\text{mm}^{2}$ shells. The subject's T1 image was also collected and processed using FreeSurfer, registered to the diffusion image and then used for tissue segmentation.

\subsubsection{Synthetic Data}
Owing to the lack of ground truth available for in vivo diffusion data, we perform quantitative evaluation of our method on synthetic data. We use the sampling scheme discussed above and consider a 2-shell acquisition with $b_{1}=1,000s/\text{mm}^{2}$ and $b_{2}=3,000s/\text{mm}^{2}$ and $M_1=M_2=M=60$, where each shell's angular samples were taken to be the corresponding gradient table from the HCP-YA subject. Gaussian noise with standard deviation $\widehat{\sigma}_{e}\approx 0.0620$, estimated from the $b=0$ images of the in vivo data as discussed Section~\ref{ssec:estimation_and_uq}, was added to the synthetic signals. A testing dataset of $N_{\text{test}} = 5,000$ samples was generated \revised{under the biologically constrained uniform priors discussed in Section~\ref{ssec:simulator}, using} Algorithm~\ref{alg:generative_model} in the supplemental materials, and used to evaluate the method's ability to recover the true unobserved model parameters.

\subsection{Computation and Implementation Details}\label{ssec:implementation_details}
{\bf Estimating fiber orientations:} We parameterize $\boldsymbol{\mathcal{A}}_{\psi}$ via the spherical U-net architecture developed in \cite{elaldi2021}.
%Briefly, the architecture consists of a set of down and upsampling paths together with skip connections along with convolutional blocks. The convolutions are defined using the rotationally equivariant graph convolution from \cite{perraudin2019}, followed by ReLU nonlinearities. The up and downsampling operations are unpooling and max pooling, respectively, defined over HEALPix grids \citep{calabretta2007}. 
While the original architecture allows for both rotationally equivariant and invariant features, we take only the former to enforce the desired rotational equivariance of the estimator. In order to ensure the antipodal symmetry of $g$, the discretized estimate $\widehat{\boldsymbol{g}}$ is projected to a high-dimensional symmetric spherical harmonic basis system (degree 20, \revised{dimension}=231) in order to preserve precise peak localization.
\par 
\noindent {\bf Inferring fiber specific model parameters:} We parameterize the mixture weights, means, and (diagonal) covariance matrices of the MDN $p_{\eta}$ using a 10-layer $L$-input channel multilayer perceptron (MLP) with ReLU non-linearities. The inputs to the MLP are the vector-valued function $\widehat{\boldsymbol{\bar{h}}}_{\mathcal{G}}^{(i)}$, discretized using a uniformly spaced grid of 100 points on $[0,1]$. We set the number of mixture components to be 3, which was selected via experimentation. While in theory the same MDN can be trained for all $n$, we find better performance by training separate models for each $n$. This is likely due to the significant distribution shift between $n$, due to the $z_{i,j}$, and hence this is important conditioning information that helps improve precision. $Q=5,000$ posterior samples are drawn for all inference tasks. The Newton's method solver implemented in the R package \textit{scam} is used for solving \eqref{eqn:SCAM}, with cubic b-spline rank $J=20$. 
\par 
For training both models, we use the Adam optimizer with a learning rate of $10^{-5}$ for 50,000 iterations and a batch size of $7,500$, with training data sampled using Algorithm~\ref{alg:marginal_generative_model} and Algorithm~\ref{alg:signal_demixx_marginal_model} of supplemental Section~\ref{ssec:sim_algos} for $\boldsymbol{\mathcal{A}}_{\psi}$ and $p_{\eta}$, respectively. All computing was performed on a Linux machine equipped with a NVIDIA RTX A6000 GPU with 48GB of RAM.

\subsection{Competing Approaches}\label{ssec:competing_approaches}
For fiber orientation estimation, we compare our method, \revised{referred to from here on as LFI (``Likelihood-Free Inverter''), in reference to the simulation-based ``likelihood-free'' training of our inversion scheme}, to two popular spherical deconvolution methods: the single-shell constrained spherical deconvolution (SS-CSD) from \cite{tournier2007}, and the multi-shell multi-tissue constrained spherical deconvolution (MS-CSD) from \cite{jeurissen2014}. We use the implementations in the python library \textit{dipy} \citep{garyfallidis2014} under the default suggested parameterizations. \revised{For SS-CSD, the response function was estimated separately for the $b=1,000 s/mm^2$ and $b=3,000s/mm^2$ shells using the dipy function \textit{auto\_response\_ssst}, applied to the $N=5,000$ test signals. Similarly, for MS-CSD, the response kernel was estimated from the same $N=5,000$ test signals using the dipy function \textit{response\_from\_mask\_msmt}, after classifying the test signals into tissue types via the \textit{TissueClassifierHMRF} class.} The number of fibers and corresponding orientations are obtained by identifying all the local maxima of the estimated ODF on a dense spherical grid using dipy's \textit{peak\_directions} function.
\par 
For biophysical parameter estimation, we compare our method to a standard approach which approximates the maximum likelihood estimator under model \eqref{eqn:standard_statistical_model} using non-linear least squares \citep{jelescu2016,harms2017}. As these methods do not have a model selection criteria, we provide the true number of fibers to the algorithm. To numerically approximate the optimization problem, we use trust reflected region optimization \citep{branch1999}. We compare two variants, one in which we randomly initialize the starting point and the optimization is run once (MLE-1), and another where multiple algorithm runs with different random initializations are used (50 in total) and the solution with the highest likelihood is taken as the final estimate (MLE-2). The latter multiple initialization scheme is a standard approach in the literature used to combat the effects of non-convexity \citep{zhang2012,panagiotaki2012}. These methods were implemented in python using the optimization routines in the \textit{scipy} library.
\par 
For the $n=1$ case, we also compare our method to a popular learning approach from \cite{reisert2017} (SMI), which estimates a regression model between the rotationally invariant features of the spherical harmonic representation of the signal and the standard model parameters~\eqref{eqn:standard_model_kernel}, under the convolution model \eqref{eqn:general_forward_model}. We use the publicly available MATLAB implementation from the \textit{Standard Model Imaging (SMI) toolbox} \citep{coelho2022}. Note that comparisons must be restricted to the $n=1$ case because this methodology does not permit the estimation of fiber-specific microstructure \citep{novikov2018}, and so the interpretation of parameters for $n\ge 2$ is not consistent between models. \revised{To ensure a fair comparison, we train SMI using the same uniform prior distributions for the kernel parameters and fix all rotational invariants order $l\ge 2$ to $1$, which is equivalent to the large $\kappa$ approximation discussed in Section~\ref{sec:models}.}

\subsection{Evaluation Metrics}
We evaluate the orientation estimation in terms of the \textit{proportion of correct peaks} (PCP), 
%defined as the proportion of samples where the correct number of fibers is estimated, 
and, conditional on correctly estimating the number of fibers, the average \textit{angular error} (AE). These quantities can be estimated using the synthetic test data via
$$
\begin{aligned}
    &\text{PCP} = \frac{1}{N_{\text{test}}}\sum_{t=1}^{N_{\text{test}}} \mathbb{I}\{\widehat{n}_{t} = n_{t}\} \\
    &\text{AE} = \frac{1}{N_{\text{test}}}\sum_{t=1}^{N_{\text{test}}}\min_{r_{1},...,r_{n_{t}}\in\{1,...,n_{t}\}:r_{i}\neq r_{j}}  \frac{1}{n_{t}}\sum_{j=1}^{n_{t}}\text{cos}^{-1}(|\widehat{\boldsymbol{m}}_{r_{j},t}^{\intercal}\boldsymbol{m}_{j,t}|)
\end{aligned}
$$
where \revised{$n_{t}$ and $\widehat{n}_t$ are the true and estimated number of fibers for the $t$'th test sample, respectively, while} $\boldsymbol{m}_{j,t}$ and $\widehat{\boldsymbol{m}}_{j,t}$ are the true and estimated orientations of the $t$'th test sample, respectively, and the minimum is required to resolve ambiguity in the labeling of the estimated orientations.
\par 
To evaluate the estimation of the biophysical parameters, we compute the median absolute error and the bias using the test set. For our method, we consider both the posterior mean (LFI-PM) and approximate MAP (LFI-MAP) point estimators discussed in Section~\ref{ssec:estimation_and_uq}. To assess the performance of the uncertainty quantification, we form the $\alpha = 0.05$ marginal HDRs and calculate two metrics using the synthetic test data. The first is the empirical coverage proportion (ECP), defined as the proportion of test samples for which the true parameter is contained in the HDR. This is a measure of the calibration of the estimated posterior, with ECP near to 0.95 indicating good calibration for this case. The second metric we compute is the average size of the HDR proportional to the total support size (HDR-S), which gives a relative measure of the ``amount'' of posterior uncertainty. 

\subsection{Results}

\subsubsection{Synthetic Data}\label{sssec:simulation_results}

\begin{table}[ht]
\centering
\begin{tabular}{lcccccc}
\hline
\multirow{2}{*}{} & \multicolumn{6}{c}{Method} \\
\cline{2-7}
 & LFI & CSD ($b=1k$) & CSD ($b=3k$) & MS-CSD & MLE-1 & MLE-2 \\
\hline
\multicolumn{6}{c}{1-Fiber} \\
\hline
PCP & 0.851	& \revised{0.667}	& \revised{0.958}	& 0.878 & -- & -- \\
AE (degrees) & 2.868 & \revised{2.925} & \revised{2.440} & 2.123 & 31.245 & 1.306 \\
\hline
\multicolumn{6}{c}{2-Fiber} \\
\hline
PCP & 0.689 & \revised{0.381} & \revised{0.319} &	0.491 & -- & -- \\
AE (degrees) & 4.484 & \revised{10.262} & \revised{6.498} & 5.336 & 34.772 & 6.824 \\
\hline
\multicolumn{6}{c}{3-Fiber} \\
\hline
PCP & 0.418 & \revised{0.338} & \revised{0.110} & 0.370 & -- & -- \\
AE (degrees) & 5.771 & \revised{18.312} & \revised{10.477} & 8.323 & 31.203 &  10.337 \\
\hline
\multicolumn{6}{c}{Inference Time} \\
\hline
Time (ms) & 0.339 & 22.887 & 22.435 & 250.935 & 784.174 & 21,171.738 \\
\hline
\end{tabular}
\caption{Synthetic data test-set averages for proportion of correct peaks (PCP), average angular errors (AE) and inference time (time) for each method.}
\label{table:angular_inf_sim}
\end{table}
\begin{figure}[!ht]
    \centering
    \includegraphics[width=\textwidth]{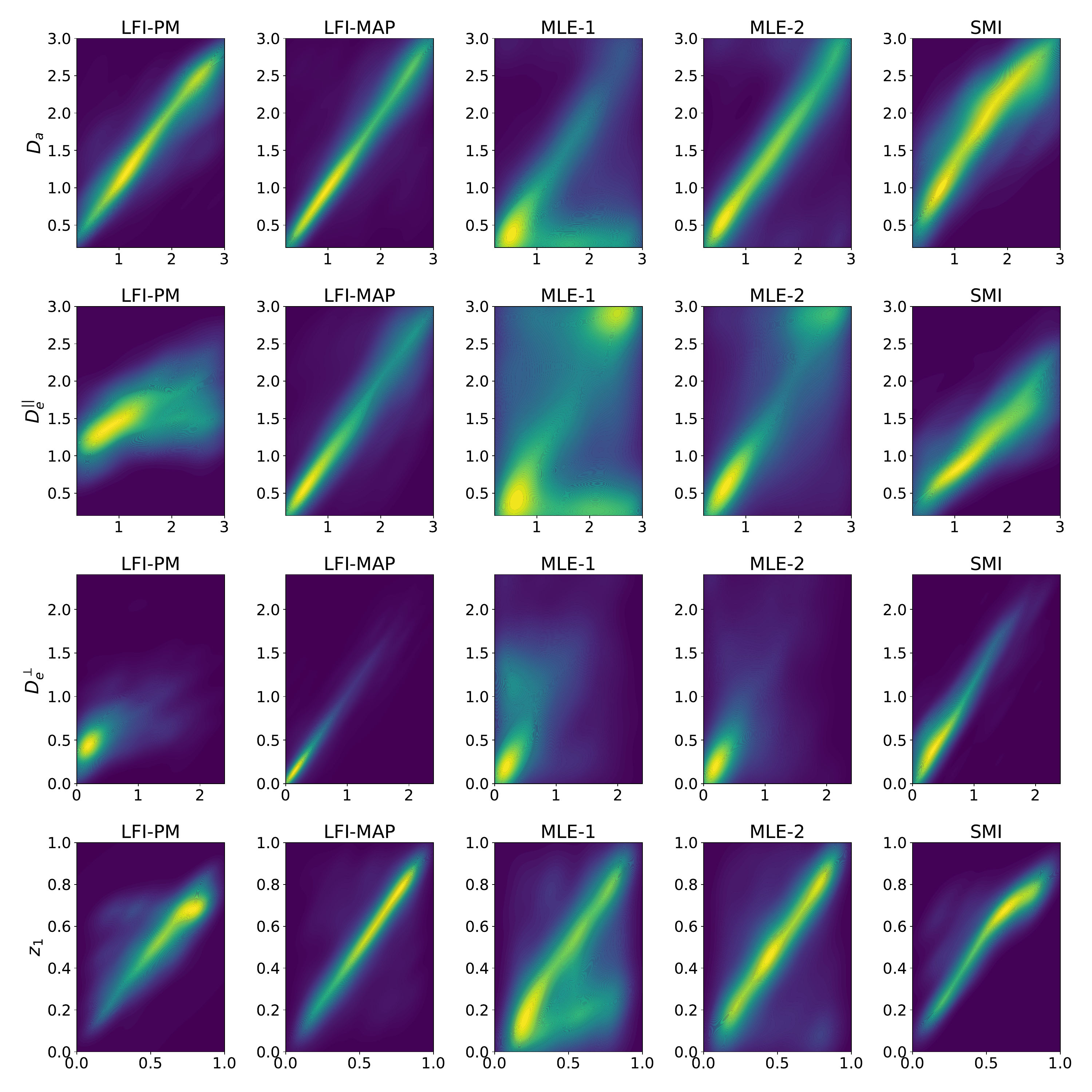}
    \caption{Density-colored scatter plots for the $n=1$ case showing the relationship between ground truth (x-axis) and estimated parameters (y-axis) for all test-set examples for each method (columns).}
    \label{fig:fib1_density_colored_scatter}
\end{figure}
\textbf{Orientational Inference}: Table~\ref{table:angular_inf_sim} shows the results for the orientation estimation. We see that our method outperforms all of the competitors in terms of both proportion of correct peaks and average angular error for both of the multi-fiber cases, while providing comparable performance to MS-CSD in the 1-fiber case. MLE-1 performs poorly for all cases, likely owing to the complexity of the loss surface. MLE-2 performs significantly better, resulting in the lowest AE for the $n=1$ case and even outperforming both the single shell CSD methods for $n > 1$. However, the MLE methods are supplied the ground truth number of fibers, and so in practice would need some additional criteria for selecting $n$. Further, MLE-2 incurs enormous computational cost, due to the requirement of repeated application of an iterative optimizer for each estimate. As our approach is amortized, we note that it is orders of magnitude faster than any of the competitors, as they all require approximating the solution to at least one optimization problem via iterative schemes. 
\par\bigskip
\noindent{\textbf{Kernel Parameter Inference}:
Table~\ref{table:shape_inf_sim} shows the average absolute error and biases for the kernel parameter estimators computed over the test set. 
Note for the multi-fiber cases, results are averaged over the fibers. For the 1-fiber case, LFI-MAP uniformly outperforms all other methods in terms of absolute error for all parameters, while also displaying low bias. \revised{LFI-PM, SMI and MLE-2, show comparable performance in terms of absolute error, with each method outperforming the others for certain parameters,  though LFI-PM demonstrates significantly lower bias for all parameters. MLE-1 demonstrates the worst performance across all parameters. To provide further context for the aggregate results summarized in Table~\ref{table:shape_inf_sim}, Figure~\ref{fig:fib1_density_colored_scatter} shows density-colored scatter plots, where the $x$ and $y$ axis are the ground truth and estimated parameters, respectively, for all test set examples in the $n=1$ case. We see that the LFI-MAP estimates for all parameters demonstrate strong positive correlation with ground truth, evident from the tight density plot around the line of equality, with little evidence of bias. Our posterior mean estimator, LFI-PM, shows some tendency for estimates to drift toward the interior of the parameter space,  particularly for $D_{e}^{||}$, resulting in slight overestimation and underestimation at the lower and upper bounds of the parameter space, respectively. SMI performs reasonably well but exhibits notable bias, which echo the aggregate results in Table~\ref{table:shape_inf_sim}, including a general underestimation of $D_{e}^{||}$ and overestimation of $D_{a}$, along with some underestimation near the boundary for $z_{1}$. MLE-1 exhibits regions of significant bias for all parameters, while MLE-2 performs reasonably well for $D_{a}$ and $z_{1}$ but has comparatively poorer performance for the extra-axonal diffusivities.}
\par 
\revised{Table~\ref{table:shape_inf_sim} shows the absolute errors of the point estimates for all methods increase with $n$, reflecting the growing complexity of the inverse problem. Notably, some parameters ($D_{a}, z_{1}$), are more reliably estimated than others ($D_{e}^{\perp},D_{e}^{||}, z_{2}$). Despite this trend, we observe that the variants of our method \revised{(LFI-MAP/LFI-PM)} are significantly more robust to the increasing dimension of the parameter space compared to the classical NLLS-based estimators \revised{(MLE-1/2)}, achieving uniformly lower average absolute errors and bias. Density colored scatter plots for the $n=2$ and $n=3$ cases are provided in Figures~\ref{fig:fib2_density_colored_scatter} and \ref{fig:fib3_density_colored_scatter} in supplemental Section~\ref{ssec:additional_synth_experiments}, along with further details and a comprehensive comparison of the estimators' performance in this regime.}
\par 
Table~\ref{table:shape_uq} shows the per-fiber ECP and HDR-S of our method. We find that the uncertainty quantification is reasonably well calibrated, with the ECP near $0.95$ for most cases. We notice that $D_{e}^{||}$ displays the largest HDR-S, indicating substantial uncertainty in the estimates. This aligns with the relatively large absolute errors for this parameter recorded in Table~\ref{table:shape_inf_sim}. These results are not surprising within the context of the poor statistical identifiability of $D_{e}^{||}$, as illustrated in Figure~\ref{fig:standard_model_geom}B. 
\begin{table}[ht!]
\centering
\begin{tabular}{ccccccc}
\hline
\multicolumn{2}{c}{\textbf{1-Fiber}} & $D_{a}$ & $D_{e}^{||}$ & $D_{e}^{\perp}$ & $z_1$ & $z_2$ \\
\hline
\multirow{2}{*}{LFI-PM} & Abs. Error & 0.199 & 0.549 & 0.275 & 0.079 & - \\
  &  Bias  &  -0.001 & 0.021 & 0.005 & 0.006 & - \\
\multirow{2}{*}{LFI-MAP} & Abs. Error & 0.133 & 0.180 & 0.077 & 0.039 & - \\
  &  Bias  &  -0.013 & 0.007 & -0.013 & 0.001 & - \\
\multirow{2}{*}{SMI} & Abs. Error & \revised{0.3485} &    \revised{0.4012}  &  \revised{0.1558} & \revised{0.0654} & - \\
  &  Bias   & \revised{-0.2435} &  \revised{0.3320} & \revised{-0.1358} & \revised{-0.0582} & - \\
\multirow{2}{*}{MLE-1} & Abs. Error & 0.482 & 0.745 & 0.424 & 0.141 & - \\
  &   Bias  & -0.598 & -0.098 & 0.258 & -0.108 & - \\
\multirow{2}{*}{MLE-2} & Abs. Error & 0.204 & 0.425 & 0.296 & 0.073 & - \\
  &   Bias  & -0.008 & 0.158 & 0.156 & -0.024 & - \\
\hline
\multicolumn{2}{c}{\textbf{2-Fiber}} & $D_{a}$ & $D_{e}^{||}$ & $D_{e}^{\perp}$ & $z_1$ & $z_2$ \\
\hline
\multirow{2}{*}{LFI-PM} & Abs. Error& 0.291 & 0.626 & 0.345 & 0.052 & 0.089 \\
&  Bias  & -0.045 & 0.035 & -0.010 & 0.018 & -0.018 \\
\multirow{2}{*}{LFI-MAP} & Abs. Error & 0.283 & 0.579 & 0.334 & 0.047 & 0.086 \\
  &  Bias  &  -0.045 & 0.017 & -0.012 & 0.002 & -0.007 \\
\multirow{2}{*}{MLE-1} & Abs. Error & 0.566 & 1.018 & 0.743 & 0.098 & 0.166 \\
& Bias & -0.327 & 0.227 & 0.513 & -0.028 & 0.136\\
\multirow{2}{*}{MLE-2} & Abs. Error & 0.355 & 0.968 & 0.721 & 0.063 & 0.136 \\
& Bias & -0.053 & 0.383 & 0.490 & 0.003 & 0.120\\
\hline
\multicolumn{2}{c}{\textbf{3-Fiber}} & $D_{a}$ & $D_{e}^{||}$ & $D_{e}^{\perp}$ & $z_1$ & $z_2$ \\
\hline
\multirow{2}{*}{LFI-PM} & Abs. Error & 0.356 & 0.671 & 0.358 & 0.037 & 0.059 \\
& Bias & -0.041 & -0.013 & 0.014 & 0.023 & -0.023 \\
\multirow{2}{*}{LFI-MAP} & Abs. Error & 0.381 & 0.684 & 0.385 & 0.040 & 0.066 \\
  &  Bias  &  -0.045 & 0.114 & 0.045 & 0.012 & -0.010 \\
\multirow{2}{*}{MLE-1} & Abs. Error & 0.670 & 1.140 & 0.878 & 0.082 & 0.102 \\
& Bias  & -0.153 & 0.265 & 0.641 & -0.013 & 0.086  \\
\multirow{2}{*}{MLE-2} & Abs. Error & 0.592 & 1.065 & 0.884 & 0.064 & 0.089  \\
& Bias  & -0.017 & 0.295 & 0.585 & 2.61e-04 & 0.075  \\
\hline
\end{tabular}
\caption{Average absolute error and bias for kernel parameter estimators for the synthetic test data. Results averaged over fibers in the multi-fiber cases.}
\label{table:shape_inf_sim}
\end{table}

\begin{table}[ht]
\centering
\begin{tabular}{ccccccc}
\hline
                   & & $D_{a}$ & $D_{e}^{||}$ & $D_{e}^{\perp}$ & $z_1$ & $z_2$      \\ \hline
\multirow{2}{*}{\textbf{1-Fiber}}  & ECP   & 0.953 & 0.941 & 0.949 & 0.953 & 0.952\\ 
    & HDR-S  & 0.466 & 0.741 & 0.513 & 0.435 & 0.435\\ \hline
\multirow{2}{*}{\textbf{2-Fiber}}   & ECP  & 0.942 & 0.919 & 0.944 & 0.959 & 0.969 \\ 
                     & HDR-S  & 0.578 & 0.804 & 0.611 & 0.269 & 0.455 \\ \hline
\multirow{2}{*}{\textbf{3-Fiber}}   & ECP  & 0.947 & 0.921 & 0.938 & 0.966 & 0.980 \\ 
                     & HDR-S  & 0.622 & 0.824 & 0.618 & 0.198 & 0.288\\ \hline
\end{tabular}
\caption{Empirical Coverage Proportion (ECP) and High Density Regions Size (HDR-S) for biophysical parameter uncertainty quantification.}
\label{table:shape_uq}
\end{table}

\subsubsection{In Vivo Data}\label{sssec:real_data_results}
\textbf{Voxel Inferences}: The top left panel of Figure~\ref{fig:per_voxel_inference} shows our method's estimate of the ODF field over a coronal slice of the HCP subject. The angular dispersion (AD, reported in radians) and detection rate (DR) for each estimated fiber at each voxel was computed using \eqref{eqn:orient_uq_measures} via $B=1,000$ bootstrapped samples from Algorithm~\ref{alg:inference_algorithm}. The top right panel of Figure~\ref{fig:per_voxel_inference} shows the AD averaged over each fiber in the voxel. We see the expected spatial pattern, with low averaged AD values within the white matter, high values in the grey matter, and intermediate values in the subcortical regions. Note that the relative lack of spatial smoothness is not surprising, as different fibers within the same voxel can exhibit different angular uncertainties, as demonstrated in the per-voxel analysis reported in the panels below.
\par 
The middle panels of Figure~\ref{fig:per_voxel_inference} show zoomed views of the colored ROIs outlined in the top left image. In each ROI, a voxel was randomly selected and the full inference results displayed in the bottom panels. The red ROI shows the characteristic crossing fiber pattern expected at the intersection of the corpus collosum (CC), corticospinal tract (CST) and superior longitudinal fascicles (SLF). Considering the inference results for the voxel encircled in red, we see that the peak belonging to the SLF (fiber 3) exhibits higher angular uncertainty, particularly as measured by the low detection rate, relative to the peaks coming from the CC (fiber 1) and CST (fiber 2). We observe further that the distributions corresponding to the intra-axonal diffusivity ($D_{a}$) and volume fraction ($z_{1}$) are shifted lower compared to fibers 1 and 2. This may suggest why the orientation of fiber 3 is more difficult to detect and localize, as these smaller values indicate less of a relative contribution to the observed signal from this fiber. 
\par 
The yellow ROI in Figure~\ref{fig:per_voxel_inference} shows the boundary of the white matter CST and subcortical region, with the encircled voxel located in the latter. We infer both fibers to have relatively small intra-axonal volume fraction, which is biologically plausible given the tissue composition in this region. The angular uncertainty, as measured by AD, is large compared to the voxels from the white matter regions, i.e. the ODFs encircled in red and blue. This may in part be due to the fact that fibers in the subcortical region tend to be less coherently aligned than the white matter.
\par
The green ROI covers part of the cortical white-grey interface, with the encircled voxel located on the boundary. The inference results suggests two fiber populations with distinct underlying character; with fiber 1 likely exhibiting larger $D_{a}$, $z_1$ and smaller $D_{e}^{\perp}$ than fiber 2. This indicates fiber 1 having more characteristics associated with whiter matter (e.g. anisotropic), with fiber 2 being more indicative of grey matter (e.g. isotropic). This is plausible given the context of the orientations, as fiber 1 appears to continue on within the white matter, whereas fiber 2 is oriented more toward the grey matter region.
\par 
The final pictured ROI (light blue) covers parts of the cingulate and corpus collosum, with the circled voxel located in the latter. The corpus collosum is known to consist of densely packed, coherently aligned unidirectional white matter fibers. These a-priori known biological features are supported in the parameter inference. Orientationally, we see a very low angular dispersion around the estimated peak, indicating a strong degree of confidence in the dominant orientation. We infer large intra-axonal diffusion $D_{a}$, and a relatively large intra-axonal volume fraction (mode near 0.5), coupled with a relatively low $D_{e}^{\perp}$, indicative of the known tissue features in this region.
\begin{figure}[ht!]
    \centering
    \includegraphics[width=\textwidth]{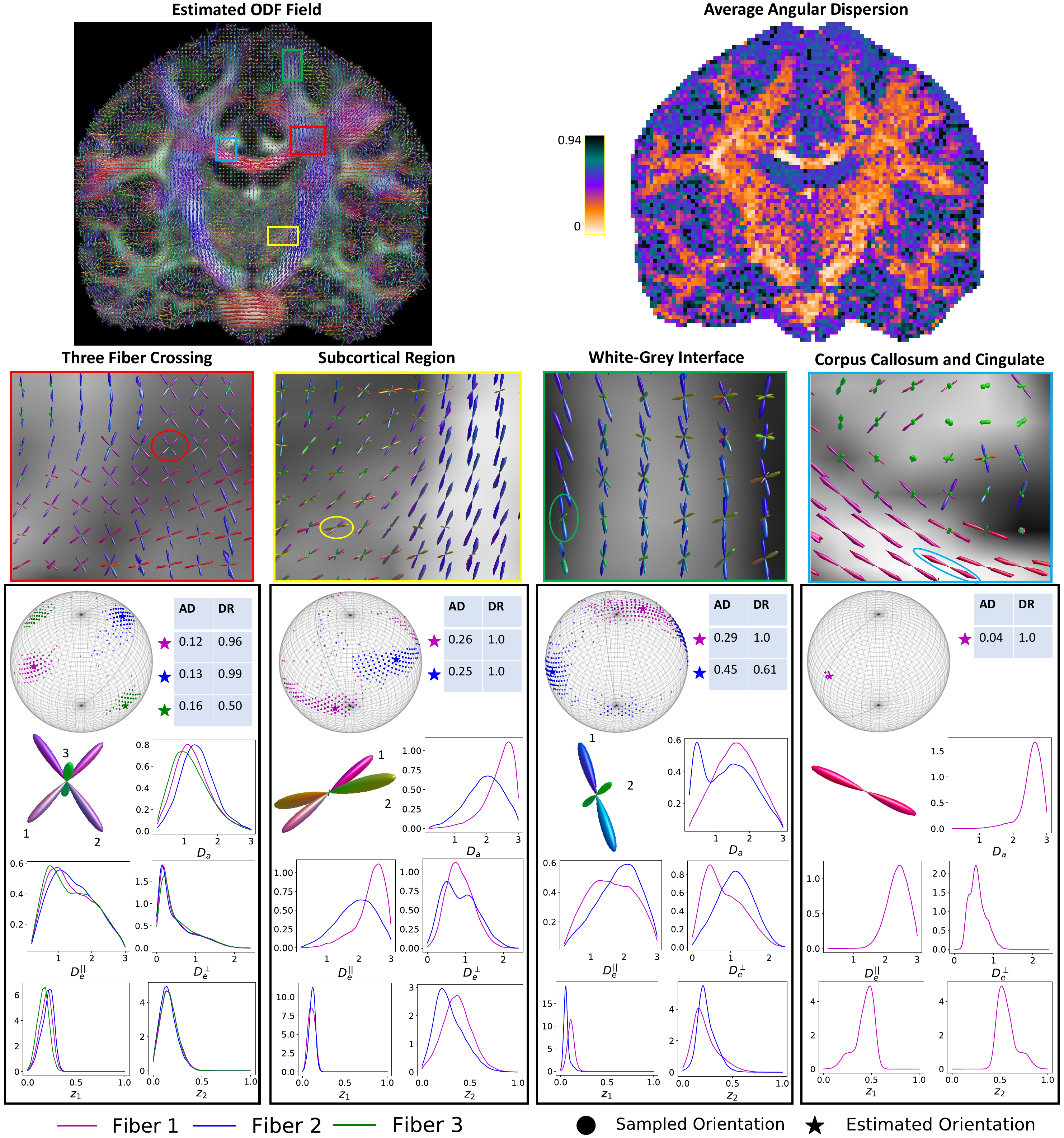}
    \caption{ODF field estimates (top left) and per-voxel average angular dispersion (AD) (top right) over a coronal cross section of the HCP-YA subject. Magnified views of the ODF field from representative ROIs are provided in the middle panels. The bottom panels show the results of the full parameter inference (posterior distributions) using Algorithm~\ref{alg:inference_algorithm} for the encircled voxel.}
    \label{fig:per_voxel_inference}
\end{figure}
\begin{figure}[ht!]
    \centering
    \includegraphics[width=\textwidth]{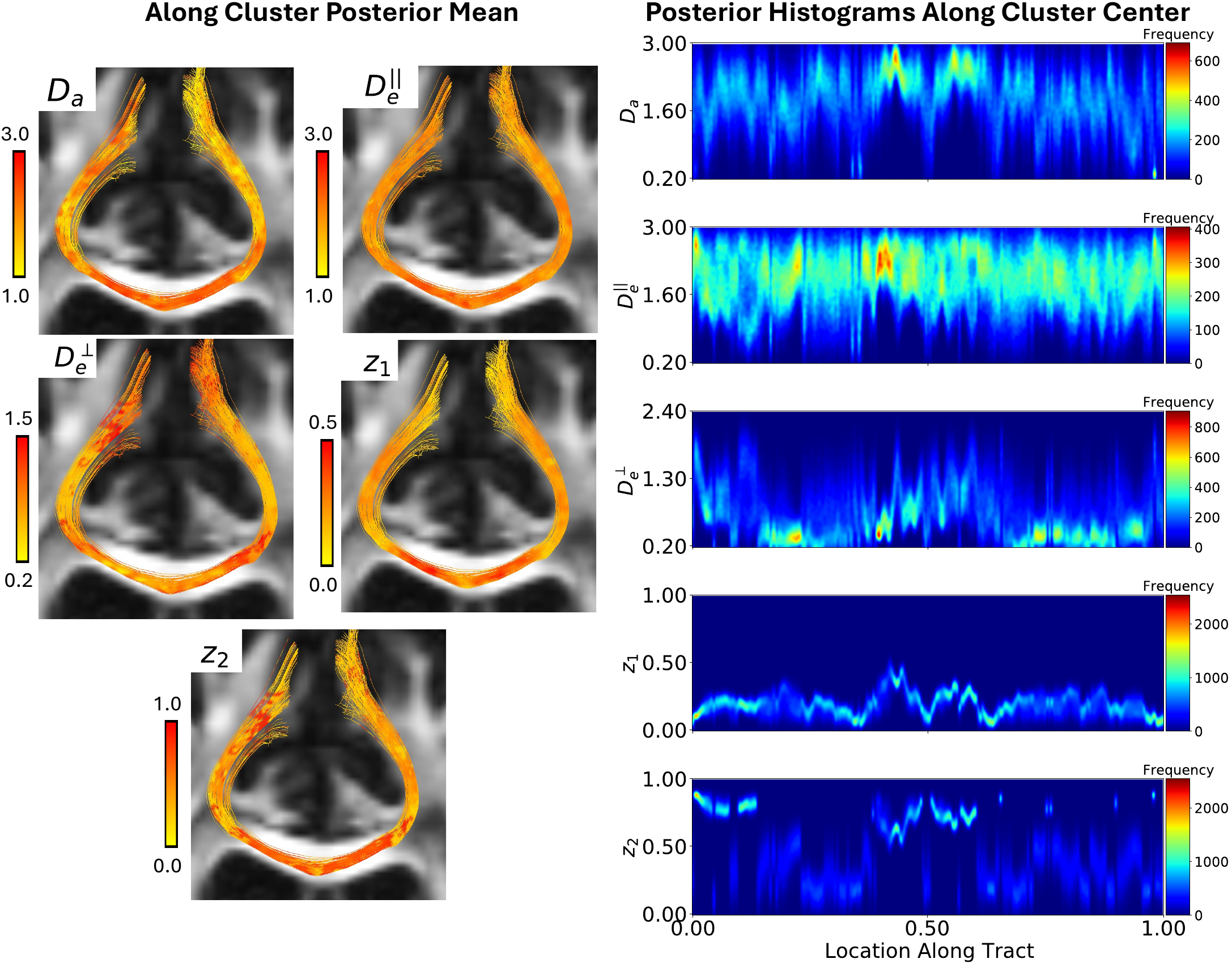}
    \caption{(Left) Fiber cluster colored by the along tract posterior mean for each kernel parameter. (Right) Histograms of posterior samples at all points along the ``most central tract'' in the cluster (as measured by the spatial curve depth).}
    \label{fig:along_tract_inference}
\end{figure}
\par\bigskip 
\textbf{Along-Tract Inferences}: To infer the underlying white matter fiber tracts, the estimated ODF field was used as input to a deterministic streamline tractography algorithm in \textit{dipy}. All white matter voxels were seeded and propagated until reaching the grey matter, with these regions defined using the FreeSurfer masks created from the subject's T1 image. The resulting streamlines were subsequently processed using the \textit{WhiteMatterAnalyis} package \citep{zhang2018}, and a sub-cluster from the corpus callosum anatomical bundle (pictured in supplemental Figure~\ref{fig:cc5_cluster}) was extracted for further analysis. The diffusion signal function along each tract was calculated using the estimated diffusion signals on the voxel grid via local linear interpolation, independently for each b-value. The interpolated functions were used to resample the signal, which was then used as input to Algorithm~\ref{alg:inference_algorithm} for estimation and inference. Due to interpolation in the tractography procedure, the estimated orientations may not exactly align with the directions implied by the tracts. For a simple workaround, we took the results from the estimated fiber most closely aligned with the local tangents along the tract as the final inference. Finally, we used the sample depth measure from \cite{micheaux2021} to rank the tracts in the cluster by spatial centrality, extracting the deepest curve, analogous to a sample median, as the most ``central tract'' for further analysis. 
\par 
The left panels of Figure~\ref{fig:along_tract_inference} show all the tracts in the cluster, colored by the estimated posterior mean for each parameter. We observe that $D_{a},z_{1},z_{2}$  are generally higher when the tracts are in the corpus collosum, which is expected due to the strong fiber alignment and lack of crossing in this region. The right panels of Figure~\ref{fig:along_tract_inference} show histograms of posterior samples at all points along the most central tract in the cluster, for each kernel parameter. The x-axis corresponds to the relative location along the tract, with 0 and 1 the starting/ending locations. The distributions along the tracts change relatively smoothly for all parameters except for $z_{2}$, where abrupt changes likely indicate regions with crossing fibers, as the extra-axonal space is now shared among multiple fibers. Echoing results from the simulated data, we notice that the diffusivity parameters tend to exhibit a greater degree of uncertainty than the volume fractions. This is particularly the case for $D_{e}^{||}$, where the posterior exhibits both high spread and several regions of sustained along-tract multi-modality.

\section{Discussion and Future Work}\label{sec:discussion}
\noindent{\textbf{Computation}: Our method was designed for  computational scalability. The full procedure in Algorithm~\ref{alg:inference_algorithm} consists of three stages, orientational inference via bootstrapping predictions of a deep inverse model, signal demixxing via solving \eqref{eqn:SCAM} and sampling the amortized posterior \eqref{eqn:appromate_posterior}. All of the steps in the first stage iteration--parametric bootstrap of \eqref{eqn:ridge_regression}, ODF estimation, and peak detection--are trivial to compute, making this resampling-based approach to uncertainty quantification scalable. Specifically, as our approach to ODF estimation is amortized, it is orders of magnitude faster than any of the competing methods recorded in Table~\ref{table:angular_inf_sim}, as they all require approximating the solution to at least one optimization problem via iterative algorithms. Replacing our estimator with any of these approaches would render this scheme computationally problematic. For signal demixxing, the synthetic data test-set average for solving \eqref{eqn:SCAM} is $\approx 0.04 s/n$, i.e. 0.04s, 0.8s, and 0.12s for the $n=1,2$ and $3$ fiber cases, respectively. The final kernel posterior sampling takes on average approximately $0.104\text{ms}/\text{sample}$. Hence, assuming an average of $n=1.5$ over $400,000$ voxels with $Q=B=100$, run in sequence, full-brain inference would take, on average, a little under $12$ hours. As the method is independent across voxels, significant reductions in computing time can be made if the processing is run in parallel.}
\par 
\noindent{\textbf{Alternative Forward and Measurement Models}: The proposed framework relies on general features of the inverse problem \eqref{eqn:general_mixture_forward_model}, e.g. rotational equivariance and additive structure, and so can be used for inference under alternative biophysical forward models and/or ODF mixture parameterizations. Replacing the Gaussian measurement model \eqref{eqn:measurement_error_model} with an alternative data likelihood can in principle be accommodated with a few minor changes to the framework, e.g. replacing the normal distribution in the parametric bootstrap step in Algorithm~\ref{alg:inference_algorithm}, as long as the distribution can be sampled from quickly. This flexibility results from the predominantly simulation-based approach to inference that was utilized, i.e. using synthetic data to train flexible deep inverse models. Another advantage of the simulation-based approach is the possibility of utilizing this framework for inference under non-analytic forward models, e.g. PDE-based \citep{fang2023}. \revised{However, as synthetic data can no longer be cheaply generated in these cases, it is likely that non-trivial augmentations will be required, presenting an interesting direction for future research. Finally, it is well known that acquiring data using high b-values, attainable using state-of-the-art scanners such as Connectome and MAGNUS, and leveraging multidimensional diffusion encoding sequences can improve the practical statistical identifiability of certain model parameters \citep{coelho2022}. Since our inference procedure is relatively agnostic to acquisition protocols, future research will focus on evaluating its performance on such modern datasets.}} 
\par 
\noindent{\textbf{Integration into Tractography Algorithms}: The along-tract inference results reported in Section~\ref{sssec:real_data_results} are somewhat ad-hoc. This is due to the piecemeal approach, where we estimate the ODF field, apply a streamline tractography algorithm, and then estimate the along tract parameters from a post-hoc local linear interpolation of the signal. This can lead to discrepancies, such as peaks inferred from the interpolated data that do not exactly align with the tangent direction along the tract. For consistent inferences, an important direction of future work is integrating the parameter estimation procedure directly into the fiber tracking.} 
\par
\noindent{\revised{\textbf{Model Misspecification}: In the specification of the forward model in Section~\ref{ssec:forward_model}, we assume $\kappa$ is very large and fixed. This assumption may be violated in complex white-matter configurations, such as fanning structures. It is well known that parameter inference under significant model misspecification—defined as systematic bias in the forward model due to missing physics that significantly impacts the observed signal—will generally lead to biased inference and overconfident estimates \citep{Brynjarsdottir_2014}. 
%Note that misspecification is different from measurement noise, which we do account for in our inference algorithm. 
Indeed, in supplemental Section~\ref{sssec:model_misspec}, we evaluate the performance of our inversion algorithm, trained under the $\kappa\rightarrow\infty$ assumption, when applied to test data where a variable $\kappa$ is included in the forward model. Results indicate that while our method maintains reasonable estimation performance and reasonably well-calibrated uncertainty quantification when the test set $\kappa$ is large (low misspecification), its performance, as expected, degrades as $\kappa$ decreases (high misspecification).}} 
\par 
\revised{To combat the deleterious effects of misspecification, one of two approaches are generally used: (1) increase the complexity of the forward model to adequately model the signal or (2) incorporate methodological additions to improve robustness of inference under model misspecification. Regarding the first approach, our framework easily accommodates modifications to the forward model by simply adapting the data generation process used to train the inverse models to include variable $\kappa$, sampled from a biologically informed prior. Preliminary results presented in supplemental Section~\ref{sssec:model_misspec} demonstrate significant improvements in both estimation accuracy and uncertainty calibration when the training data incorporates $\kappa$, compared to the previous case where the model is trained without $\kappa$ but tested on data where variable $\kappa$ is included. However, we recognize that models incorporating $\kappa$ are surely not fully well specified in any meaningful sense. Therefore, rather than attempting to invert increasingly complex and high-dimensional forward models, we suggest that pursuing the second approach of improving robustness to misspecification within the amortized inference setting (pretraining models to approximate posterior quantities) holds greater promise. This is currently an active area of research \citep{huang2023learning,schmitt2024detectingmodelmisspecificationamortized}, and the present use case serves as a strong motivation for further algorithmic development along these lines.}

\section{Conclusion}\label{sec:conclusion}
In this work, we propose a novel method for biophysical parameter estimation and uncertainty quantification in general mixture deconvolution models of brain microstructure using diffusion MRI data. Instead of inferring all model parameters simultaneously, our approach uses a series of deep neural network-based inverse models, each tailored to different marginal and conditional subproblems and trained through simulations. The inverse modules are then applied sequentially to observed imaging data for full parameter inference. In simulation studies, we show that the proposed method generally outperforms standard competing approaches, both in terms of lower errors and computational times, while also providing well-calibrated uncertainties that can be computed rapidly. Qualitative analysis of our method using in vivo diffusion data from the Human Connectome Project shows promising results in accurately capturing fiber specific microstructural properties.

%% file: supplement_revision.tex
\section{Proof of Theorem~\ref{thm:rotational_equivaraince}}
\label{ssec:rot_equivariane_proof}
\begin{proof}
Let $\boldsymbol{P}_{V} := (\boldsymbol{p}_1, ..., \boldsymbol{p}_V)\subset\mathbb{S}^2$ be a high resolution spherical grid. Since the signal estimates are formed independently between b-values, without loss of generality, we assume $L=1$ and drop the subscript $L$ notation for clarity. Given rotation matrix $\boldsymbol{R}\in\mathbb{SO}(3)$, we define the rotations 
$$
\begin{aligned}
    &\boldsymbol{g} = (g(\boldsymbol{p}_1), ...,g(\boldsymbol{p}_V)), \qquad T_{\boldsymbol{R}}\left[\boldsymbol{g}\right] = (g(\boldsymbol{R}^{-1}\boldsymbol{p}_1), ..., g(\boldsymbol{R}^{-1}\boldsymbol{p}_V)) \\
    &\widehat{\boldsymbol{f}} = (\widehat{f}(\boldsymbol{p}_1), ...,\widehat{f}(\boldsymbol{p}_V)), \qquad T_{\boldsymbol{R}}\left[\widehat{\boldsymbol{f}}\right] = (\widehat{f}(\boldsymbol{R}^{-1}\boldsymbol{p}_1), ..., \widehat{f}(\boldsymbol{R}^{-1}\boldsymbol{p}_V)) \\
    &\boldsymbol{\mathcal{A}}(\widehat{\boldsymbol{f}}) = (\mathcal{A}(\widehat{\boldsymbol{f}})(\boldsymbol{p}_1), ...,\mathcal{A}(\widehat{\boldsymbol{f}})(\boldsymbol{p}_V)), \qquad T_{\boldsymbol{R}}\left[\boldsymbol{\mathcal{A}}(\widehat{\boldsymbol{f}})\right] = (\mathcal{A}(\widehat{\boldsymbol{f}})(\boldsymbol{R}^{-1}\boldsymbol{p}_1), ..., \mathcal{A}(\widehat{\boldsymbol{f}})(\boldsymbol{R}^{-1}\boldsymbol{p}_V)),
\end{aligned}
$$
and $T_{\boldsymbol{R}}^{-1}$ is the inverse transformation defined similarly. To show the desired rotational equivariance, we must show that 
$$
T_{\boldsymbol{R}}\boldsymbol{\widehat{\mathcal{A}}}(\widehat{\boldsymbol{f}}) = \boldsymbol{\widehat{\mathcal{A}}}(T_{\boldsymbol{R}}\left[\widehat{\boldsymbol{f}}\right]).
$$
holds.
\par 
First, we note that the prior of $p(\boldsymbol{g})$ is rotationally invariant, i.e. $p(\boldsymbol{g}) = p(T_{\boldsymbol{R}}\left[\boldsymbol{g}\right])$. To see this notice that, for any $n$, the joint prior distribution of the $\boldsymbol{m}^{(n)}$ outlined in Section~\ref{ssec:simulator} only depends on the angle between them. Clearly, these angles are preserved under rotation, and hence the rotational invariance of the prior follows. 
\par 
Assuming no regularization, it is easy to see that the distribution of the harmonic coefficient estimator is given by:
$$
    \widehat{\boldsymbol{c}} \sim \mathcal{N}(\boldsymbol{c}, \sigma_{e}^2\left[\boldsymbol{\Phi}_V^{\intercal} \boldsymbol{\Phi}_V\right]^{-1}),
$$
where $\boldsymbol{c}$ is the true harmonic coefficient vector and $\boldsymbol{\Phi}_V\in\mathbb{R}^{V\times K}$ is the spherical harmonic basis evaluation matrix over the grid $\boldsymbol{P}_{V}$. For simplicity, we assume $K$ is taken large enough to assume a negligible truncation bias (i.e., signal is approximately band-limited). We assume that $V$ is large enough so that $\boldsymbol{\Phi}_V^{\intercal} \boldsymbol{\Phi}_V \approx V\boldsymbol{I}_{K}$, which follows due to the orthogonality of the harmonics, since for large $V$ the $(k,k^{\prime})$ element of $V^{-1}\boldsymbol{\Phi}_V^{\intercal} \boldsymbol{\Phi}_V$ is 
$$
\frac{1}{V}\sum_{v=1}^V\phi_{k}(\boldsymbol{p}_v)\phi_{k^{\prime}}(\boldsymbol{p}_v) \approx \int_{\mathbb{S}^2}\phi_{k}(\boldsymbol{p})\phi_{k^{\prime}}(\boldsymbol{p})d\boldsymbol{p} = \mathbb{I}\{k=k^{\prime}\}.
$$
These assumption imply the following (approximate) function space distribution of the signal estimator:
\begin{equation}\label{eqn:basis_function_estimator_distr}
    \widehat{\boldsymbol{f}} \sim \mathcal{N}\left((f(\boldsymbol{p}_1), ..., f(\boldsymbol{p}_V))^{\intercal}, \frac{\sigma^2_{e}}{V}\boldsymbol{\Phi}_V\boldsymbol{\Phi}_V^{\intercal}\right).    
\end{equation}
Let $\tilde{\boldsymbol{u}}=\boldsymbol{R}^{-1}\boldsymbol{u}$ and observe the following identity for the conditional mean:
\begin{equation}\label{eqn:rot_invariant_mean}
\begin{aligned}
    &\mathbb{E}\left[\widehat{f}(\boldsymbol{p}) | \boldsymbol{\xi}^{(n)}=\boldsymbol{\xi}^{(n)}, n=n, g=T_{\boldsymbol{R}}\left[g\right]\right] = \int_{\mathbb{S}^2}\sum_{i=1}^n h_{\mathcal{G}}(\boldsymbol{p}^{\intercal}\boldsymbol{u}|b,\boldsymbol{\xi}_i)g(\boldsymbol{R}^{-1}\boldsymbol{u}|\boldsymbol{m}_i)d\boldsymbol{u}\\ &=  \int_{\mathbb{S}^2}\sum_{i=1}^n h_{\mathcal{G}}(\boldsymbol{p}^{\intercal}\boldsymbol{R}\tilde{\boldsymbol{u}}|b,\boldsymbol{\xi}_i)g(\tilde{\boldsymbol{u}}|\boldsymbol{m}_i)d\tilde{\boldsymbol{u}} \\
    &=  \int_{\mathbb{S}^2}\sum_{i=1}^n h_{\mathcal{G}}((\boldsymbol{R}^{-1}\boldsymbol{p})^{\intercal}\tilde{\boldsymbol{u}}|b,\boldsymbol{\xi}_i)g(\tilde{\boldsymbol{u}}|\boldsymbol{m}_i)d\tilde{\boldsymbol{u}} = f(\boldsymbol{R}^{-1}\boldsymbol{p}) \\
    &= \mathbb{E}\left[\widehat{f}(\boldsymbol{R}^{-1}\boldsymbol{p}) | \boldsymbol{\xi}^{(n)}=\boldsymbol{\xi}^{(n)}, n=n, g=g\right].
\end{aligned}
\end{equation}
Now, consider the covariance matrix of \eqref{eqn:basis_function_estimator_distr}, whose elements are of the form $[\boldsymbol{\Phi}_{V}\boldsymbol{\Phi}_{V}^\intercal]_{vv^{\prime}} = \sum_{k=1}^K\phi_k(\boldsymbol{p}_v)\phi_k(\boldsymbol{p}_{v^{\prime}})$. We must investigate the properties of this sum. Using traditional notation, denote the set of real spherical harmonics of degree $l$ as: $\mathcal{H}_{l}:=\{Y_{-l}^l, ..., Y_{l}^l\}$ (see \citeSupp{descoteaux2007} equation 3 for precise definition). The real symmetric basis $\{\phi_{1},...,\phi_{K}\}$ is formed from a relabeling of $\bigcup_{l=0}^{R}\mathcal{H}_{2l}$, i.e., only even degrees are taken. Then, by the spherical harmonic addition theorem, we have that 
$$
\sum_{k=1}^K\phi_k(\boldsymbol{p}_v)\phi_k(\boldsymbol{p}_{v^{\prime}}) = \sum_{l=0}^R\sum_{m=-21}^{2l}Y_{m}^{2l}(\boldsymbol{p}_v)Y_{m}^{2l}(\boldsymbol{p}_{v^{\prime}}) \propto P_{2l}(\langle \boldsymbol{p}_v, \boldsymbol{p}_{v^{\prime}}\rangle),
$$
where $P_{l}$ is the Legendre polynomial of degee $l$.
Crucially, this sum is rotationally invariant, which is inherited from the invariance of the inner product: 
$\langle \boldsymbol{p}_v, \boldsymbol{p}_{v^{\prime}}\rangle = \langle \boldsymbol{R}^{-1}\boldsymbol{p}_v, \boldsymbol{R}^{-1}\boldsymbol{p}_{v^{\prime}}\rangle$. Hence, $\boldsymbol{\Phi}_{V}\boldsymbol{\Phi}_{V}^{\intercal}$ is invariant with respect to any rotation of the grid $\boldsymbol{P}_{V}$. Coupling this property with \eqref{eqn:rot_invariant_mean}, it follows that the vectorized distribution $\boldsymbol{\boldsymbol{f}}$ is rotationally equivariant in the sense
$$
p(\widehat{\boldsymbol{f}}|\boldsymbol{\xi}^{(n)}, n, T_{\boldsymbol{R}}\left[\boldsymbol{g}\right]) = p(T_{\boldsymbol{R}}\left[\widehat{\boldsymbol{f}}\right]|\boldsymbol{\xi}^{(n)}, n, \boldsymbol{g}). 
$$
Integrating both sides w.r.t. $p(\boldsymbol{\xi}^{(n)}, n)$, this in turn implies the marginal condition 
$$
p(\widehat{\boldsymbol{f}}|T_{\boldsymbol{R}}\left[\boldsymbol{g}\right]) = p(T_{\boldsymbol{R}}\left[\widehat{\boldsymbol{f}}\right]|\boldsymbol{g}).
$$ 
Now, given that 
$$
\begin{aligned}
    \frac{1}{V}\|T_{\boldsymbol{R}}\left[\boldsymbol{g}\right] - T_{\boldsymbol{R}}\left[\boldsymbol{\mathcal{A}}(\widehat{\boldsymbol{f}})\right]\|_{2}^2 &\approx \int_{\mathbb{S}^2}(g(\boldsymbol{R}^{-1}\boldsymbol{p}) -  \mathcal{A}(\widehat{\boldsymbol{f}})(\boldsymbol{R}^{-1}\boldsymbol{p}))^2d\boldsymbol{p} \\ 
    &= \int_{\mathbb{S}^2}(g(\boldsymbol{p}) -  \mathcal{A}(\widehat{\boldsymbol{f}})(\boldsymbol{p}))^2d\boldsymbol{p} \\
    &\approx \frac{1}{V}\|\boldsymbol{g} - \boldsymbol{\mathcal{A}}(\widehat{\boldsymbol{f}})\|_{2}^2,
\end{aligned}
$$
we have that the $\|\cdot\|_{2}$ norm is approximately invariant to $T_{\boldsymbol{R}}$.
\par
Putting this all together, we have 
\begin{equation}\label{eqn:rotated_risk}
    \begin{aligned}
        &\int \int \|\boldsymbol{g} - \boldsymbol{\mathcal{A}}(\widehat{\boldsymbol{f}})\|_2^2p(\widehat{\boldsymbol{f}}|\boldsymbol{g})p(\boldsymbol{g} )d\widehat{\boldsymbol{f}}d\boldsymbol{g} \\
        &\overset{\text{prior invariance}}{=} \int \int \|T_{\boldsymbol{R}}\left[\boldsymbol{g}\right] - \boldsymbol{\mathcal{A}}(\widehat{\boldsymbol{f}})\|_2^2p(\widehat{\boldsymbol{f}}|T_{\boldsymbol{R}}\left[\boldsymbol{g}\right] )p(\boldsymbol{g})d\widehat{\boldsymbol{f}}d\boldsymbol{g} \\
         &\overset{\text{model equivariance}}{=} \int \int \|T_{\boldsymbol{R}}\left[\boldsymbol{g}\right] - \boldsymbol{\mathcal{A}}(T_{\boldsymbol{R}}\left[\widehat{\boldsymbol{f}}\right])\|_2^2p(T_{\boldsymbol{R}}\left[\widehat{\boldsymbol{f}}\right]|\boldsymbol{g})p(\boldsymbol{g})d\widehat{\boldsymbol{f}}d\boldsymbol{g}\\
          &\overset{\text{norm invariance}}{=} \int \int \|\boldsymbol{g} - T_{\boldsymbol{R}}^{-1}\boldsymbol{\mathcal{A}}(T_{\boldsymbol{R}}\left[\widehat{\boldsymbol{f}}\right])\|_2^2p(T_{\boldsymbol{R}}\left[\widehat{\boldsymbol{f}}\right]|\boldsymbol{g})p(\boldsymbol{g})d\widehat{\boldsymbol{f}}d\boldsymbol{g}
    \end{aligned}
\end{equation}
Hence, if $\boldsymbol{\widehat{\mathcal{A}}}$ minimizes  \eqref{eqn:bayes_risk}, it must also minimize 
$$
    \int \int \|\boldsymbol{g} - T_{\boldsymbol{R}}^{-1}\boldsymbol{\mathcal{A}}(T_{\boldsymbol{R}}\left[\widehat{\boldsymbol{f}}\right])\|_2^2p(T_{\boldsymbol{R}}\left[\widehat{\boldsymbol{f}}\right]|\boldsymbol{g})p(\boldsymbol{g})d\widehat{\boldsymbol{f}}d\boldsymbol{g}.
$$
which in turn implies the risk minimizer must also satisfy
$$
\widehat{\boldsymbol{\mathcal{A}}}(\widehat{\boldsymbol{f}}) = T_{\boldsymbol{R}}^{-1}\widehat{\boldsymbol{\mathcal{A}}}(T_{\boldsymbol{R}}\left[\widehat{\boldsymbol{f}}\right]),
$$
which gives the equivariance condition
$$
T_{\boldsymbol{R}}\widehat{\boldsymbol{\mathcal{A}}}(\widehat{\boldsymbol{f}}) = \widehat{\boldsymbol{\mathcal{A}}}(T_{\boldsymbol{R}}\left[\widehat{\boldsymbol{f}}\right])
$$
as desired.
\end{proof}

\section{Simulation Algorithms}\label{ssec:sim_algos}
Algorithm~\ref{alg:generative_model} provides pseudo-code for sampling from the joint distribution of diffusion signals and model parameters. Algorithm~\ref{alg:marginal_generative_model} provides pseudo-code for sampling from the joint distribution of diffusion signal function estimates and ODF. Algorithm~\ref{alg:signal_demixx_marginal_model} provides pseudo-code for sampling from the joint distribution of biophysical kernel parameters and estimated demixxed signal curves. 
\begin{algorithm}[ht!]
\caption{Sample from $p(\boldsymbol{S}, \boldsymbol{\xi}^{(n)}, \boldsymbol{m}^{(n)})$}
\label{alg:generative_model}
\begin{algorithmic}[1]
    \STATE \textbf{Input}: Sampling design $\boldsymbol{P}_{M,l}$, $b_l$ for $l=1,...,L$, measurement error variance $\sigma_{e}^2$
    \STATE Sample $n \sim \text{Unif}\left([1,...,n_{max}]\right)$
    \STATE Sample $(z_{1,1}, z_{1,2}, z_{2,1}, ..., z_{n,1}, z_{n,2})$ using rejection sampling with a proposal Dirichlet$((\frac{1}{2n}, ...,\frac{1}{2n}))$ and rejection criteria $z_{i,1} < 0.1$ for any $i=1,...,n$.
    \STATE For $i=1,...,n$, independently sample $(D_{a,i}, D_{e,i}^{||}, D_{e,i}^{\perp})\sim \text{Unif}(\Xi)$ using rejection sampling with a box-uniform proposal, rejecting any samples that don't respect the polytope constraints.
    \STATE Form $\boldsymbol{\xi}_i = (D_{a,i}, D_{e,i}^{||}, D_{e,i}^{\perp}, z_{i,1}, z_{i,2})$ for $i=1,...,n$ 
    \STATE Sample $(\boldsymbol{m}_1, ..., \boldsymbol{m}_n)$ using rejection sampling with a proposal $\prod_{i=1}^n\text{Unif}(\mathbb{S}^2_{+})$ and rejection criteria defined by the crossing angle constraints outlined in 
    %Section~\ref{ssec:sim_algos}
    \revised{Section~\ref{ssec:simulator}}
    \STATE Simulate $\boldsymbol{S}=(\boldsymbol{s}_1, ..., \boldsymbol{s}_L)$ from \eqref{eqn:standard_statistical_model}      
    \RETURN $(\boldsymbol{S}, \boldsymbol{\xi}^{(n)},\boldsymbol{m}^{(n)})$
\end{algorithmic}
\end{algorithm}

\begin{algorithm}[ht!]
\caption{Sample from $p(\widehat{\boldsymbol{f}}_L, \boldsymbol{g})$}
\label{alg:marginal_generative_model}
\begin{algorithmic}[1]
    \STATE \textbf{Input}: High-resolution spherical grid points $\boldsymbol{P}_{V}=\{\boldsymbol{p}_1, ..., \boldsymbol{p}_{V}\}$.
    \STATE Sample $(\boldsymbol{S}, \boldsymbol{\xi}^{(n)},\boldsymbol{m}^{(n)})$ via Algorithm~\ref{alg:generative_model}
    \STATE Compute $\widehat{f}_L$ via \eqref{eqn:ridge_regression} using simulated $\boldsymbol{S}$
    \STATE Set $\boldsymbol{g} = \frac{1}{n}\sum_{i=1}^n g_i(\boldsymbol{p}_v|\boldsymbol{m}_i)$, $v=1,...,V$
    \STATE Set $\widehat{\boldsymbol{f}}_L$ as the evaluations of $\widehat{f}_L$ over $\boldsymbol{P}_{V}$ 
    \RETURN $(\widehat{\boldsymbol{f}}_L, \boldsymbol{g})$
\end{algorithmic}
\end{algorithm}
\begin{algorithm}[ht!]
\caption{Sample from $p(\boldsymbol{\xi}_i,\widehat{\boldsymbol{\bar{h}}}_{\mathcal{G}}^{(i)}|n)$}
\label{alg:signal_demixx_marginal_model}
\begin{algorithmic}[1]
    \STATE Sample $(\boldsymbol{S}, \boldsymbol{\xi}^{(n)},\boldsymbol{m}^{(n)})$ via Algorithm~\ref{alg:generative_model}
    \STATE Compute $\boldsymbol{t}_{l,m}^{*} = ((\boldsymbol{p}_{l,m}^\intercal\boldsymbol{m}_1)^2, ..., (\boldsymbol{p}_{l,m}^\intercal\boldsymbol{m}_n)^2)$ and $\boldsymbol{\Gamma}_{mj}^{(i,l)} = \gamma_j(t^{*}_{l,m,i})$, for $i=1,...,n$, $l=1,...,L$, $m=1,...,M$
    \STATE Estimate $\widehat{\boldsymbol{a}}^{(1,l)},...,\widehat{\boldsymbol{a}}^{(n,l)}$ via \eqref{eqn:SCAM} for $l=1,...,L$
    \STATE Form $\widehat{\boldsymbol{\bar{h}}}_{\mathcal{G}}^{(i)}(t_i^{*}) = (\boldsymbol{\gamma}(t_{i}^{*})^{\intercal}\widehat{\boldsymbol{a}}^{(i,1)}, ..., \boldsymbol{\gamma}(t_{i}^{*})^{\intercal}\widehat{\boldsymbol{a}}^{(i,L)})$
    \STATE Return $\{\widehat{\boldsymbol{\bar{h}}}_{\mathcal{G}}^{(i)}, \boldsymbol{\xi}_i, n\}$, $i=1,...,n$
\end{algorithmic}
\end{algorithm}
\section{Additional Methodological Details}\label{sec:supp_method_details}
%Figure~\ref{fig:method_overview} provides a schematic overview of the proposed methodology.
\subsection{Signal Demixxing}\label{ssec:GAM_signal_demix}
\subsubsection{Monotonic Spline Parameterization}\label{ssec:demix_algo}
We note that the  functions $\bar{h}_{\mathcal{G}}^{(i,l)}$'s are monotonic decreasing in $t_{i}^{*}$. In order to enforce this property in our estimation, we parameterize them using basis expansion over monotonic decreasing cubic b-splines of rank $J$, constructed by following the formulation in \citeSupp{pya2015}. For completeness, we provide salient details on the construction of this spline model, and refer the interested reader to the reference for a more thorough treatment. Define $\boldsymbol{\gamma}(t) = (\gamma_1(t), ...,\gamma_J(t))$ to be the rank $J$ cubic b-spline basis. We model $\bar{h}_{\mathcal{G}}^{(i,l)}(t_{i}^{*}) \approx \boldsymbol{\gamma}(t_{i}^{*})^{\intercal}\boldsymbol{a}^{(i,j)}$, where $\boldsymbol{a}^{(i,j)} := \boldsymbol{\Sigma}\tilde{\boldsymbol{\beta}}^{(i,j)}$,  $\tilde{\boldsymbol{\beta}}^{(i,l)} = (\beta_1^{(i,l)}, \exp(\beta_2^{(i,l)}),...,\exp(\beta_J^{(i,l)}))$ with $\boldsymbol{\beta}^{(i,l)} =(\beta_1^{(i,l)}, ...,\beta_J^{(i,l)})$, and $\boldsymbol{\Sigma}\in\mathbb{R}^{J\times J}$ defined element-wise as 
$$
\boldsymbol{\Sigma}_{jj^\prime} = \begin{cases}
    0 \text{ for } j < j^{\prime} \\
    1 \text{ for } j^{\prime}=1, j\ge 1\\
    -1 \text{ for } j^{\prime}\ge 2, j\ge j^{\prime}.
\end{cases}
$$
It can be validated that modeling the basis expansion coefficients in this way guarantees the desired monotonicity property in the estimates. The optimization problem \eqref{eqn:SCAM} can be written in the new parameterization as follows:
\begin{equation}\label{eqn:SCAM_fullparam}
\begin{aligned}
   \widehat{\mu}_{l},\widehat{\boldsymbol{\beta}}^{(1,l)},...,\widehat{\boldsymbol{\beta}}^{(n,l)} = &\min_{\mu_{l},\boldsymbol{\beta}^{(1,l)},...,\boldsymbol{\beta}^{(n,l)}}\left\|\boldsymbol{s}_{l} - \left[\boldsymbol{1}_{M}\mu_{l} + \sum_{i=1}^n\boldsymbol{\Gamma}^{(i)}\boldsymbol{\Sigma}\tilde{\boldsymbol{\beta}}^{(i,l)}\right]\right\|_2^2 \\
   &\text{s.t.}  \quad  \boldsymbol{1}_{M}^{\intercal}\boldsymbol{\Gamma}^{(i)}\boldsymbol{\Sigma}\tilde{\boldsymbol{\beta}}^{(i,l)} = 0, \text{ for }i = 1, ..., n,
\end{aligned}
\end{equation}
and (approximately) solved using the Newton's method solver proposed in \citeSupp{pya2015}. 
\subsubsection{A Brief Note on Statistical Properties}
\revised{Recall that the signal demixing approach outlined in Section~\ref{ssec:signal_demix} relies on a transformation $\boldsymbol{p}\mapsto \boldsymbol{t}^{*}_n$, mapping the domain (at each b-shell) $\mathbb{S}^2\mapsto[0,1]^n$. This transformation is designed to exploit the \textit{additive structure} of the signal function in the transformed space. As shown in Figure~\ref{fig:signal_demixxing} of the main text for $n=2$, the signal function in this transformed space is a surface over the $n$-dimensional hyper-cube which is additive in the marginal decay curves. This additive structure provides a way of demixing the signal. Specifically, the kernel parameters of the $i$'th fiber can be inferred using only the $i$'th estimated decay curve, under approximate conditional independence.} 
\par 
\revised{Optimal estimates of the additive signal curves are known to enjoy good statistical properties, as the statistical errors $\|\widehat{\bar{\boldsymbol{h}}}_{\mathcal{G}}^{(i)}-\bar{\boldsymbol{h}}_{\mathcal{G}}^{(i)}\|$ are (asymptotically in number of samples $M$) independent of $n$ \citepSupp{Stone1985,Horowitz2006}. This implies that the demixxed signal functions can be estimated with a (asymptotically) constant statistical rate regardless of $n$, which is a result of the special additive structure and is not the case for general $n$-dimensional functions (which require exponentially more data as $n$ increases). This structure offers a type of protection against the curse of dimensionality for larger $n$.} 
\par 
\revised{Although the convergence of signal demixing-based inference was not explicitly investigated in this work, as $M$ was fixed for all experiments, we suggest this as an interesting direction for future research. Specifically, it would be valuable to evaluate whether the theoretically fast statistical convergence of the decay curve estimates for large $n$ translates into practically faster convergence rates for fiber-specific microstructure compared to, for example, joint optimization in the original space.}

\subsection{Computing Marginal HDRs}\label{ssec:HDR_computation}
We form HDRs marginally for each parameter in $\xi_{i,p}\in \boldsymbol{\xi}_i$, as follows. First, we bin the support of $\xi_{i,p}$ into $B$ disjoint intervals. We then draw $Q$ samples from the amortized posterior $\boldsymbol{\xi}_{i,q} \sim p_{\widehat{\eta}}(\boldsymbol{\xi}_{i}|\widehat{\boldsymbol{\bar{h}}}_{\mathcal{G}}^{(i)},n)$ and, for each binned sub-interval, we calculate the proportion of samples $\xi_{i,p,q}$ belonging to each bin. The bins are sorted in descending order according to this proportion and the smallest index $r$ in the sorted list such that the cumulative sum of the first $r$ elements is greater than  $(1-\alpha)$ is identified. Finally, the HDR is defined as the union of the bins corresponding to the first $r$ elements of the sorted list. Algorithm~\ref{alg:HDR} provides pseudo-code for this procedure.
\begin{algorithm}
\caption{Forming Highest Density Regions (HDRs) for each parameter}
\label{alg:HDR}
\begin{algorithmic}[1]
\STATE \textbf{Input}: Number of bins $B$, number of samples $Q$, the level $\alpha$
\STATE  Draw $Q$ samples from the amortized posterior: $\boldsymbol{\xi}_{i,q} \sim p_{\widehat{\eta}}(\boldsymbol{\xi}_{i}|\widehat{\boldsymbol{\bar{h}}}_{\mathcal{G}}^{(i)},n)$
\FOR{each parameter $\xi_{i,p} \in \boldsymbol{\xi}_i$}
    \STATE Bin the support of $\xi_{i,p}$ into $B$ disjoint intervals
    \STATE Initialize array \textit{counts} of size $B$ to zero
    \FOR{$q=1,...,Q$}
        \STATE Identify the bin index $k$ for $\xi_{i,p,q}$
        \STATE \textit{counts}[$k$] $ = $ \textit{counts}[$k$] + 1
    \ENDFOR
    \STATE \textit{proportions}[$k$] $ = $ \textit{counts}[$k$]/$Q$ 
    \STATE \textit{sorted\_proportions} = \text{sort}(\textit{proportions}, order=descending)
    \STATE Find the smallest index $r$ such that cumulative sum of \textit{sorted\_proportions} up to $r$ is greater than $(1-\alpha)$
    \STATE Define HDR as the union of the bins corresponding to the first $r$ indices in \textit{sorted\_proportions}
\ENDFOR
\STATE \textbf{Output}: Return HDR
\end{algorithmic}
\end{algorithm}

\section{Additional Experimental Results}\label{sec:additional_experiments}
\subsection{Synthetic Data}\label{ssec:additional_synth_experiments}
\begin{figure}[!ht]
    \centering
    \includegraphics[width=\textwidth]{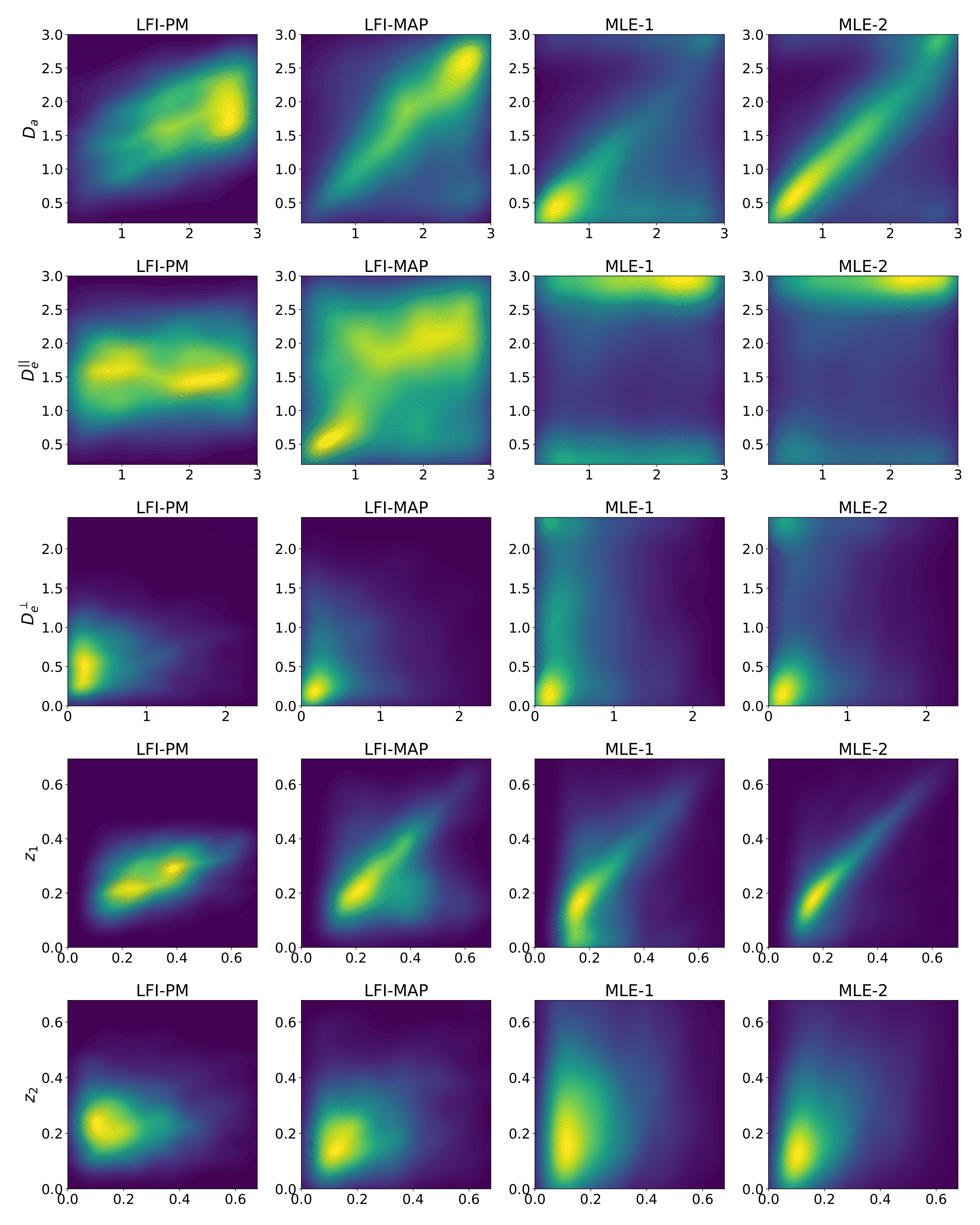}
    \caption{Density-colored scatter plots for the $n=2$ case showing the relationship between ground truth (x-axis) and estimated parameters (y-axis) for all test-set examples for each method (columns).}
    \label{fig:fib2_density_colored_scatter}
\end{figure}
\begin{figure}[!ht]
    \centering
    \includegraphics[width=\textwidth]{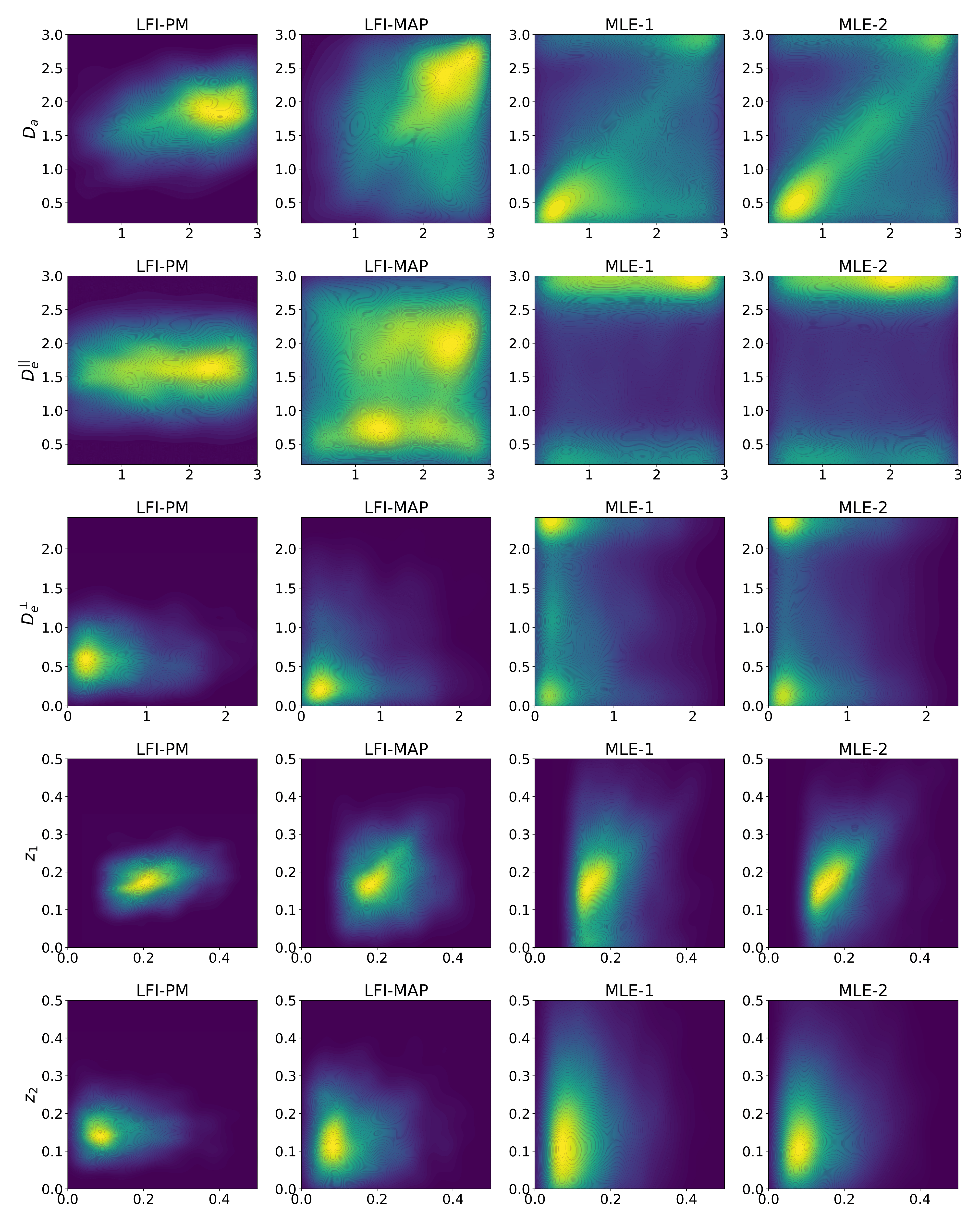}
    \caption{Density-colored scatter plots for the $n=3$ case showing the relationship between ground truth (x-axis) and estimated parameters (y-axis) for all test-set examples for each method (columns).}
    \label{fig:fib3_density_colored_scatter}
\end{figure}
\subsubsection{Additional Details on Estimation for $n\ge 2$}
\revised{Figures~\ref{fig:fib2_density_colored_scatter} and \ref{fig:fib3_density_colored_scatter} show density colored scatter plots for estimated versus ground truth parameters for all estimators. Consistent with the aggregate results shown in Table~\ref{table:shape_inf_sim} of the main text, for the $n=2$ fiber case, we observe a general increase in estimation errors across all parameters for all estimators compared to the $n=1$ results. This is reflected in weaker correlations between predictions and ground truth, with density plots appearing more diffuse and/or somewhat tilted relative to the line of equality. For $D_{a}$, all estimators exhibit positive correlation with ground truth. However, MLE 1/2 and LFI-MAP all display some examples of significant under estimation near the upper bound of ground truth (non-trivial density in bottom right), while MLE 1/2 also exhibit substantial overestimation near the lower bound of ground truth (non-trivial density in upper left). This behavior likely arises from increased multi-modality in this higher-dimensional case versus $n=1$. In contrast, the $D_{a}$ estimates of LFI-PM do not exhibit such extreme errors at either boundary. Instead, we observe a relatively minor ``tendency toward the interior'' behavior (as noted for $D_{e}^{||}$ in the $n=1$ case in the main text), with modest over- and underestimation at the boundaries compared to the other estimators. We speculate this is due to the averaging effect of the posterior mean computation mitigating extreme errors that result from selecting an inaccurate mode, and could be desirable behavior in this regime. For $D_{e}^{||}$, LFI-MAP is the only estimator producing somewhat reasonable estimates, though the density plot appears more diffuse relative to the line of equality compared to the $n=1$ case, reflecting the increased difficulty of the inverse problem. LFI-MAP and LFI-PM perform reasonably well for small ground truth values of $D_{e}^{\perp}$, but estimation errors increase as the ground truth values increase. The MLE 1/2 estimators both exhibit regions of dramatic overestimation (mode in the top left of Figure~\ref{fig:fib2_density_colored_scatter}). All methods do reasonably well for intra-axonal volume fraction, while LFI-MAP appears to have the best performance for $z_{2}$, which is backed up by the aggregate results in manuscript Table~\ref{table:shape_inf_sim}.} 
\par 
\revised{For the $n=3$ fiber case, we observe a further increase in estimation errors across all parameters and estimators compared to the $n=2$ fiber case, consistent with the aggregate results presented in Table~\ref{table:shape_inf_sim} in the main text. The comparison of estimators for $D_{a}$ is similar to the $n=2$ case, except for the general trend of increased errors across all estimators. In this regime, estimates of $D_{e}^{||}$ or $D_{e}^{\perp}$ from any estimator should be treated with caution, as they are highly unreliable. On the other hand, the methods perform reasonably well for the intra-axonal volume fraction $z_1$, although MLE 1/2 appear to exhibit greater bias for low values. These findings suggest that, while the signal is very weakly informative for the extra-axonal parameters, it may still be sufficiently informative to allow for usable estimates for the intra-axonal volume fraction and gives some limited information for $D_{a}$. This could be beneficial in certain scenarios, such as maintaining accurate tractography in crossing fiber situations, where matching intra-axonal volume fractions could help identify and select the appropriate peak.}
\subsubsection{Comparison With Trivial Prior Mean Estimator}
\revised{The biologically constrained uniform prior over model parameters outlined in Section~\ref{ssec:simulator} was chosen deliberately as a non-informative prior to avoid biasing inference. In the case when the signal provides little to no information about the parameters of interest, we would expect our approximate Bayesian posterior to essentially return the prior. Hence, we can define the ``trivial estimator'' as simply taking the prior mean. This estimator is unbiased, but minimally informative in the sense that it is not using any information in the signal to update its estimate (just returning the prior). As such, it serves as a type of baseline representing the simplest possible approach. Table~\ref{table:prior_mean_error} shows the average absolute errors of the prior mean estimator over the test set. Comparing these to the results reported in Table~\ref{table:shape_inf_sim} of the main text, we make the following observations: First, our inversion algorithm does not merely return the mean of the training distribution for any of the $n=1,2$ or $3$-fiber cases. Second, our inversion algorithm shows a better relative improvement over the trivial prior mean estimator for the volume fractions ($z_1/z_2$) and $D_{a}$, than it does for the extra-axonal diffusivities, $D_{e}^{||},D_{e}^{\perp}$. This is reflected in the model uncertainty estimates in Table~\ref{table:shape_uq}, where the high-density regions of $D_{e}^{||}$ (and to a lesser extent $D_{e}^{\perp}$) take up a relatively larger portion of the parameter range than the other parameters (quantified by HDR-S), indicating a more spread out posterior distribution. Finally, the prior mean appears to ``outperform'' some of the competing methods in terms of average error. However, this advantage arises not from containing more information but rather from the substantial bias present in the other estimators.}
\begin{table}[ht]
\centering
\begin{tabular}{ccccccc}
\hline
                   & & $D_{a}$ & $D_{e}^{||}$ & $D_{e}^{\perp}$ & $z_1$ & $z_2$      \\ \hline
\textbf{1-Fiber}  & Abs. Error   & 0.703 & 0.700 & 0.399 & 0.197 & -- \\ 
\textbf{2-Fiber}   & Abs. Error  & 0.710 & 0.697 & 0.397 & 0.113 & 0.116 \\ 
\textbf{3-Fiber}   & Abs. Error  & 0.706 & 0.701 & 0.400 & 0.073 & 0.074 \\ 
\end{tabular}
\caption{Average absolute error results for the naive prior mean estimator on the synthetic test set.}
\label{table:prior_mean_error}
\end{table}
\subsubsection{Inference Under Forward Model Misspecification}\label{sssec:model_misspec}

\begin{table}[ht]
\centering
\begin{tabular}{ccccccc}
\hline
                   & Metric & $D_{a}$ & $D_{e}^{||}$ & $D_{e}^{\perp}$ & $z$ \\ \hline
\multicolumn{6}{c}{\textbf{$\kappa \rightarrow \infty$}} \\ \hline
\multirow{2}{*}{LFI-PM}  & Abs. Error & 0.199 & 0.549 & 0.275 & 0.079 \\ 
                         & Bias       & -0.001 & 0.021 & 0.005 & 0.006 \\ \hline
\multirow{2}{*}{LFI-MAP} & Abs. Error & 0.133 & 0.180 & 0.077 & 0.039 \\ 
                         & Bias       & -0.013 & 0.007 & -0.013 & 0.001 \\ \hline
\textbf{ECP}           &  &     0.953       &   0.941     &    0.949   &    0.953        \\ \hline
\multicolumn{6}{c}{\textbf{$\kappa = 30$}} \\ \hline
\multirow{2}{*}{LFI-PM}  & Abs. Error  & 0.296 & 0.588 & 0.279 & 0.138 \\ 
                         & Bias            &   -0.241    &   0.069    &   -0.023    &  -0.082     \\ \hline
\multirow{2}{*}{LFI-MAP} & Abs. Error  & 0.299 & 0.317 & 0.117 & 0.07 \\ 
                         & Bias            &  -0.352     &    0.408   &   0.083    &  -0.026     \\ \hline
\textbf{ECP}             &                 & 0.935 & 0.889 & 0.901 & 0.854 \\ \hline
\multicolumn{6}{c}{\textbf{$\kappa = 20$}} \\ \hline
\multirow{2}{*}{LFI-PM}  & Abs. Error & 0.374 & 0.570 & 0.270 & 0.155 \\ 
                         & Bias            &   -0.345    &    0.103   &   -0.054    &   -0.107    \\ \hline
\multirow{2}{*}{LFI-MAP} & Abs. Error  & 0.396 & 0.315 & 0.108 & 0.087 \\ 
                         & Bias            &    -0.449   &    0.374   &   0.003    &    -0.062   \\ \hline
\textbf{ECP}             &                 & 0.904 & 0.861 & 0.866 & 0.786 \\ \hline
\multicolumn{6}{c}{\textbf{$\kappa = 10$}} \\ \hline
\multirow{2}{*}{LFI-PM}  & Abs. Error  & 0.607 & 0.640 & 0.279 & 0.200 \\ 
                         & Bias            &  -0.620     &   0.178    &   -0.029    &   -0.157    \\ \hline
\multirow{2}{*}{LFI-MAP} & Abs. Error  & 0.678 & 0.409 & 0.136 & 0.140 \\ 
                         & Bias            &  -0.714     &  0.405     &   -0.012    &  -0.101     \\ \hline
\textbf{ECP}             &                 & 0.715 & 0.799 & 0.773 & 0.622 \\ \hline
\end{tabular}
\caption{Biophysical parameter inference for different levels of misspecification ($\kappa$ values) for the $n=1$ case.}
\label{table:mispec_results_combined_nfib_1}
\end{table}

\begin{table}[ht]
\centering
\begin{tabular}{ccccccc}
\hline
                   & & $D_{a}$ & $D_{e}^{||}$ & $D_{e}^{\perp}$ & $z_1$ & $z_2$ \\ \hline
\multicolumn{7}{c}{$(\kappa_1,\kappa_2) \rightarrow \infty$}  \\ \hline
\multirow{2}{*}{LFI-PM} & Abs. Error& 0.291 & 0.626 & 0.345 & 0.052 & 0.089 \\
&  Bias  & -0.045 & 0.035 & -0.010 & 0.018 & -0.018 \\
\multirow{2}{*}{LFI-MAP} & Abs. Error & 0.283 & 0.579 & 0.334 & 0.047 & 0.086 \\
  &  Bias  &  -0.045 & 0.017 & -0.012 & 0.002 & -0.007 \\\hline          
\multicolumn{7}{c}{\textbf{$(\kappa_1,\kappa_2) \sim \text{Unif}([25,35]^2)$}} \\ \hline
\multirow{2}{*}{LFI-PM}  & Abs. Error & 0.372 & 0.617 & 0.353 & 0.061 & 0.113 \\ 
                                      & Bias      & 0.146 & -0.021 & -0.020 & 0.041 & -0.051 \\ \hline
\multirow{2}{*}{LFI-MAP} & Abs. Error & 0.401 & 0.625 & 0.341 & 0.058 & 0.093 \\ 
                                      & Bias      & 0.184 & -0.096 & 0.012 & 0.034 & -0.023 \\ \hline
\textbf{ECP}                          &                 & 0.938 & 0.913 & 0.947 & 0.943 & 0.961 \\ \hline
\multicolumn{7}{c}{\textbf{$(\kappa_1,\kappa_2) \sim \text{Unif}([10,15]^2)$}} \\ \hline
\multirow{2}{*}{LFI-PM} & Abs. Error & 0.503 & 0.622 & 0.356 & 0.073 & 0.129 \\ 
                                      & Bias      & 0.314 & -0.024 & -0.023 & 0.062 & -0.073 \\ \hline
\multirow{2}{*}{LFI-MAP} & Abs. Error & 0.590 & 0.663 & 0.335 & 0.074 & 0.106 \\ 
                                      & Bias      & 0.420 & -0.147 & 0.007 & 0.050 & -0.049 \\ \hline
\textbf{ECP}                          &                 & 0.886 & 0.889 & 0.925 & 0.838 & 0.921 \\ \hline
\end{tabular}
\caption{Biophysical parameter inference under model misspecification for different $\kappa$ ranges for the $n=2$ case.}
\label{table:mispec_results_combined}
\end{table}

\begin{table}
    \centering
    \caption{Inference results for $\kappa\sim\text{Unif}([5, 35])$ in the model.}
    \begin{tabular}{cccccc}
        \hline 
        & &  $D_{a}$ & $D_{e}^{||}$ & $D_{e}^{\perp}$ & $z_1$ \\ 
        \hline
        \multirow{2}{*}{PM} & Abs. Errors & 0.3383 & 0.6432 & 0.3172 & 0.1370 \\ \cline{2-6}
        & Bias & -0.0087 & -0.0071 & -0.0135 & -0.0067 \\ \hline
        \multirow{2}{*}{MAP} & Abs. Errors & 0.3295 & 0.4029 & 0.1823 & 0.1287 \\ \cline{2-6}
        & Bias & 0.1500 & -0.0860 & 0.0053 & 0.0164  \\ \hline
    \textbf{ECP}        & & 0.946 &  0.9304 & 0.9454 & 0.9658  \\ \hline    
    \end{tabular}
    \label{tab:parameter_errors_bias}
\end{table}

\revised{To examine the performance of our model under varying levels of model misspecification, we apply our pre-trained inverter (trained using the forward model from Section~\ref{ssec:forward_model} assuming $\kappa$ very large and fixed) to test data generated with variable $\kappa$. That is, we model the noiseless test signals as }
\begin{equation}\label{eqn:signal_model_wkappa}
    \begin{aligned}
    f(\boldsymbol{p}, b) \approx \sum_{i=1}^n \int_{\mathbb{S}^2}&\Big[z_{i,1}\exp\left(-b D_{i,a}(\boldsymbol{p}^{\intercal}\boldsymbol{u})^2\right) + 
    z_{i,2}\exp\left(-b D_{i,e}^{\perp} - b(D_{i,e}^{||} - D_{i,e}^{\perp})(\boldsymbol{p}^{\intercal}\boldsymbol{u})^2\right)\Big]\times \\
    &C(\kappa_i)\text{exp}(\kappa_i (\boldsymbol{m}_i^{\intercal}\boldsymbol{u})^2)d\boldsymbol{u}.
\end{aligned}
\end{equation}

\revised{Table~\ref{table:mispec_results_combined_nfib_1} shows the performance of both LFI-PM and LFI-MAP estimators, along with uncertainty calibration as measured by the ECP of the $95\%$ posterior HDR, for the $n=1$ fiber case with $\kappa=30,20,10$, representing increasing levels of misspecification. For convenience, the high-concentration test data results (\(\kappa \to \infty\)) are also included, copied from Table~\ref{table:shape_inf_sim} of the main manuscript. As expected, we see that the performance degrades with decreasing $\kappa$, with reasonable results shown for $\kappa = 30$ (and to a lesser extent $\kappa = 20$), while inference for $\kappa=10$ is quite problematic in both estimation and uncertainty quantification for $D_{a}$ and $D_{e}^{||}$. }
\par 
\revised{Similarly, Table~\ref{table:mispec_results_combined} shows the results for misspecified inference for the $n=2$ case obtained by applying the pre-trained inverter to out of distribution test data from both $(\kappa_1,\kappa_2)\sim\text{Unif}([25,35]^2)$ and $(\kappa_1,\kappa_2)\sim\text{Unif}([10,15]^2)$. The $\kappa_1,\kappa_2\rightarrow\infty$ (in-distribution) high-concentration 2-fiber test data results from main text Table~\ref{table:shape_inf_sim} are also provided for comparison. We observe that the uncertainty remains quite well calibrated for $(\kappa_1,\kappa_2)\sim\text{Unif}([25,35]^2)$, with all of the ECP relatively close to the nominal $0.95$ level. For $(\kappa_1,\kappa_2)\sim\text{Unif}([10,15]^2)$, the uncertainty is generally underestimated for all parameters, with some reporting only mild over-confidence ($D_{e}^{\perp}$, $z_1$), and others significantly under-covered ($z_1$). As was observed in the $n=1$ results, the degradation in estimation performance, measured by absolute errors and bias, is significantly more pronounced in the lower $\kappa$ case. This outcome is expected, as the lower $\kappa$ represents a greater deviation from the forward model ($\kappa\rightarrow\infty$) used to train the posterior estimator. The parameter estimate which suffers the most significant degradation (for both posterior mean and MAP estimators) is intra-axonal volume fraction $D_{a}$, both in terms of absolute errors and bias.
This could be because both parameters have similar effects on the shape of the signals, i.e., both higher $D_{a}$ and $\kappa$ will lead to more anisotropy.}
\par 
\revised{One approach for dealing with (known) model misspecification is to incorporate the additional complexity into the forward model and perform inference on this expanded parameter space. 
The main contribution of our work is the two-stage approach to parameter inference, specifically designed to exploit the general structure of forward models like those in Equation~\eqref{eqn:general_mixture_forward_model}, rather than being restricted to the particular form chosen in Section~\ref{sec:models}. Therefore, we can easily accommodate inclusion of $\kappa$ into our framework by replacing step 7 in Algorithm~\ref{alg:generative_model} and instead generating  noiseless signals from \eqref{eqn:signal_model_wkappa} under the chosen acquisition design and then adding Gaussian noise. To demonstrate this flexibility, Table~\ref{tab:parameter_errors_bias} provides some preliminary results for the $n=1$ case when $\kappa$ is included in the forward model under the same HCP-like acquisition scheme. The biological prior $\text{Unif}([5, 35])$ is used, and the remainder of the inversion procedure is kept the same. Given these results are averaged over the test set sampled from $\kappa\sim \text{Unif}([5, 35])$, we see significant improvement over the corresponding misspecified inference results reported in Table~\ref{table:mispec_results_combined_nfib_1} in terms of errors, bias and, importantly, uncertainty quantification, as now the ECP is back near $0.95$ for all parameters. When comparing the results of the two estimators, we observed that LFI-MAP generally has lower absolute errors but increased bias compared to LFI-PM.  Results from \citepSupp{jelescu2016} indicate that the forward model with standard model kernel and Watson density with variable $\kappa$ exhibits a substantial degree of ill-posedness. This manifests in a multi-modal posterior (and likelihood) with at least two plausible solutions: one tending towards \( D_a < D_{e}^{||} \), low \( z_{1} \), and high $\kappa$, and the other with \( D_a > D_{e}^{||} \), high $z_{1}$, and low value of $\kappa$. Given this context, the lower-error but higher-bias performance of LFI-MAP likely arises from its tendency to select the ``wrong mode'' for some non-trivial number of test cases. In contrast, LFI-PM averages over the modes, leading to higher errors (as it lands between the modes) but lower bias. These estimators have different properties, and it is likely application dependent which one (if either) a user would want to report. Nonetheless, we emphasize the flexibility of our methodology as a general-purpose solver, capable of adapting to different forward models (and acquisition schemes).} 

\subsection{Real Data}
\begin{figure}[ht!]
    \centering
    \includegraphics[scale=0.75]{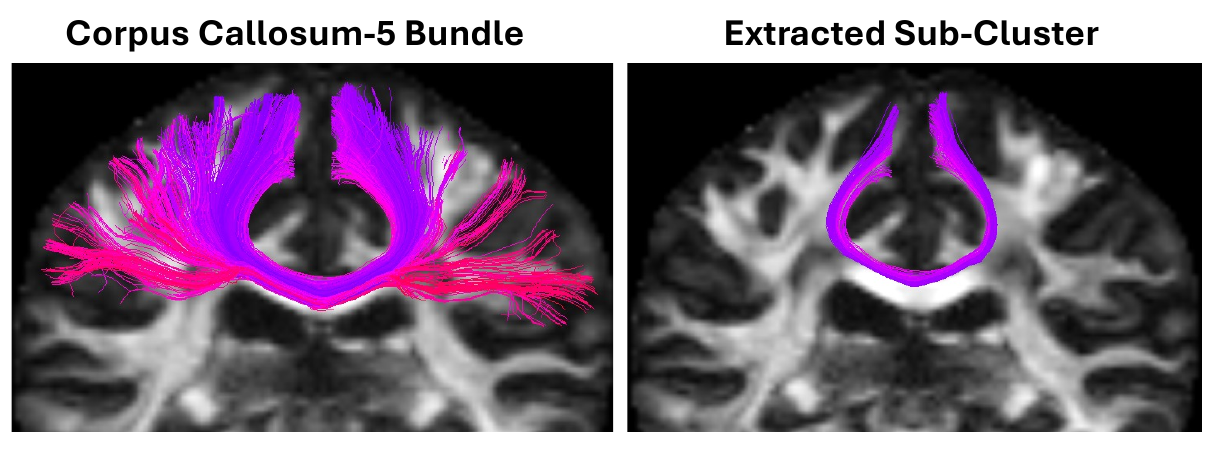}
    \caption{(Left) Corpus Callosum-5 (CC5) anatomical bundle obtained from streamline tractography clustering results using \textit{WhiteMatterAnaylsis} package. (Right) Sub-cluster extracted from CC5 bundle.}
    \label{fig:cc5_cluster}
\end{figure}

Figure~\ref{fig:cc5_cluster} shows a sub-cluster of part of the corpus callosum bundle extracted from the tractography results using \textit{WhiteMatterAnaylsis}.
